\documentclass[useAMS,usenatbib]{mnras}
\usepackage{graphicx}
\usepackage{amsmath}
\usepackage{amssymb}
\usepackage{subfigure}
\usepackage{url}
\usepackage{multirow}

\newcommand{\kpc}{\rm\thinspace kpc}

\newcommand{\km}{\rm\thinspace km}

\newcommand{\cm}{\rm\thinspace cm}

\newcommand{\s}{\rm\thinspace s}
\newcommand{\ks}{\rm\thinspace ks}

\newcommand{\Hz}{\rm\thinspace Hz}

\newcommand{\Msun}{\hbox{$\rm\thinspace M_{\odot}$}}

\newcommand{\keV}{\rm\thinspace keV}

\newcommand{\erg}{\rm\thinspace erg}

\newcommand{\ergps}{\hbox{$\erg\s^{-1}\,$}}

\newcommand{\cts}{\rm\thinspace ct}
\newcommand{\ctsps}{\hbox{$\cts\s^{-1}\,$}}

\newcommand{\ergcmps}{\hbox{$\erg\cm\ps\,$}}

\newcommand{\kmps}{\hbox{$\km\s^{-1}\,$}}

\newcommand{\ps}{\hbox{$\s^{-1}\,$}}

\newcommand{\rg}{\rm\thinspace $r_\mathrm{g}$}

\title[The corona of I\,Zw\,1]{Revealing structure and evolution within the corona of the Seyfert galaxy I\,Zw\,1}
\author[D. R. Wilkins et al.]{D. R. Wilkins$^1$\thanks{E-mail: dan.wilkins@stanford.edu}\thanks{Einstein Fellow}, L. C. Gallo$^2$, C. V. Silva$^{3,4}$, E. Costantini$^{3,4}$, W. N. Brandt$^{5,6,7}$ \newauthor and G. A. Kriss$^8$\\
$^1$Kavli Institute for Particle Astrophysics and Cosmology, Stanford University, 452 Lomita Mall, Stanford, CA 94305, USA \\
$^2$Department of Astronomy \& Physics, Saint Mary's University, Halifax, NS. B3H 3C3, Canada \\
$^3$SRON, Netherlands Institute for Space Research, Sorbonnelaan 2, 3584 CA Utrecht, The Netherlands \\
$^4$Anton Pannekoeck Institute for Astronomy, University of Amsterdam, Science Park 904, 1098 XH Amsterdam, The Netherlands \\
$^5$Department of Astronomy and Astrophysics, 525 Davey Lab, The Pennsylvania State University, University Park, PA 16802, USA \\
$^6$Institute for Gravitation and the Cosmos, The Pennsylvania State University, University Park, PA 16802, USA \\
$^7$Department of Physics, 104 Davey Lab, The Pennsylvania State University, University Park, PA 16802, USA \\
$^8$Space Telescope Science Institute, 3700 San Martin Drive, Baltimore, MD 21218, USA
}
\begin{document}

\date{Accepted 2017 July 17. Received 2017 July 12; in original form 2017 January 30}

\pagerange{\pageref{firstpage}--\pageref{lastpage}} \pubyear{2017}

\maketitle

\label{firstpage}

\begin{abstract}
X-ray spectral timing analysis is presented of \textit{XMM-Newton} observations of the narrow line Seyfert 1 galaxy I Zwicky 1 (I\,Zw\,1) taken in 2015 January. After exploring the effect of background flaring on timing analyses, X-ray time lags between the reflection-dominated 0.3-1.0\keV\ energy and continuum-dominated 1.0-4.0\keV\ band are measured, indicative of reverberation off the inner accretion disc. The reverberation lag time is seen to vary as a step function in frequency; across lower frequency components of the variability, $3\times 10^{-4}$ to $1.2\times 10^{-3}$\Hz\ a lag of 160\s\ is measured, but the lag shortens to $(59\pm 4)$\s\ above $1.2\times 10^{-3}$\Hz. The lag-energy spectrum reveals differing profiles between these ranges with a change in the dip showing the earliest arriving photons. The low frequency signal indicates reverberation of X-rays emitted from a corona extended at low height over the disc while at high frequencies, variability is generated in a collimated core of the corona through which luminosity fluctuations propagate upwards. Principal component analysis of the variability supports this interpretation, showing uncorrelated variation in the spectral slope of two power law continuum components. The distinct evolution of the two components of the corona is seen as a flare passes inwards from the extended to the collimated portion. An increase in variability in the extended corona was found preceding the initial increase in X-ray flux. Variability from the extended corona was seen to die away as the flare passed into the collimated core leading to a second sharper increase in the X-ray count rate.
\end{abstract}



\section{Introduction}
Active galactic nuclei (AGN) are some of the most luminous objects in the Universe, powered by the accretion of material onto a supermassive black hole in the centre of a galaxy. In particular, intense X-ray continuum emission with a power law spectrum, extending in many cases to over 100\keV\ before a cut-off, is produced from the inner regions of the accretion flow; the putative corona. 

Although it is responsible for a significant fraction of the energy output from the AGN, the nature of the corona remains a mystery, both in terms of its structure and how energy is injected into it from the accretion flow. It is largely considered to be an energetic population of particles, accelerated by magnetic fields that are associated with the accretion disc and black hole magnetosphere, that produce the X-ray continuum through the Comptonisation of thermal photons, predominantly ultraviolet, that are emitted from the accretion disc.

In recent years, a significant amount has been learned about the structure of the corona from the reflection and reverberation of the continuum emission it produces off of the accretion disc. When a power law X-ray continuum is incident upon the geometrically thin, optically thick accretion disc of relatively cold gas, reprocessing through Compton scattering, photoelectric absorption and subsequent fluorescent line emission and thermal bremsstrahlung produces a characteristic `reflection' spectrum \citep{ross_fabian}. This spectrum is shifted in energy and blurred by relativistic effects in the disc; Doppler shifts due to the orbital motion of disc material and gravitational redshifts due to the proximity of the reflecting material to the black hole. Most notably, this produces the prominent iron K$\alpha$ fluorescence line at 6.4\keV, which exhibits an extended redshifted wing \citep{fabian+89} as well as a `soft excess' of emission below 1\keV\ as relativistic blurring causes numerous emission lines due to iron, oxygen and nitrogen, among other elements, to be blended together.

Since the Doppler shift and gravitational redshift vary as a function of radius in the accretion disc, \citet{1h0707_emis_paper} show that fitting the broadened iron K$\alpha$ fluorescence line in time-averaged spectra from long observations reveals the illumination pattern, or emissivity profile, of the accretion disc, which in turn reveals the geometry of the corona that illuminates it \citep{understanding_emis_paper}. In a number of sources, the corona is found to extend over the inner few tens of gravitational radii of the accretion disc (1\rg\,$=GM/c^2$ is the characteristic scale-length in the gravitational field, with the event horizon of a maximally rotating Kerr black hole at 1\rg).

Due to the additional light travel time between the corona and the reflecting accretion disc, reverberation time lags are seen between correlated variability in the directly observed continuum emission and the reflected emission. The discovery of these time lags (\citealt{fabian+09,reverb_review} for a review) has added a further dimension to the study of AGN coronae. Short reverberation lags show that while extending radially over the disc, coronae extend, at most, one or two gravitational radii vertically above the disc \citep{lag_spectra_paper,cackett_ngc4151}. X-ray reverberation is seen in the high frequency components of the variability, corresponding to the short light travel timescales to the inner disc and measuring the time lag between X-rays of different energies over this frequency range (the \textit{lag-energy spectrum}) reveals a profile reminiscent of the reflection spectrum \citep{kara_1h0707}. Delayed response is seen in X-rays below 1\keV, corresponding to the broad soft X-ray excess reflected from the accretion disc, as well as in the iron K$\alpha$ fluorescence line. Redshifted photons in the low energy wing of the line are seen to respond sooner due to the shorter light travel time from the corona to the inner regions of the disc from where these redshifted photons arise, compared to the core of the line produced from the outer disc \citep{zoghbi+2012}. A consistent lag-energy profile from 2 to 10\keV\ is seen across a sample of Seyfert galaxies \citep{kara+13} supporting the interpretation of reflection and reverberation from the inner regions of the accretion disc.

On the other hand, at low frequencies, a steadily increasing time lag is seen from lower to higher energy X-rays. Such a time lag cannot be explained by reverberation. This `hard lag' is a well known feature in X-ray binaries \citep{miyamoto+88,miyamoto+89,nowak+99} and is often attributed to the propagation of fluctuations in mass accretion rate through the disc, energising the less energetic outer parts of the corona before reaching the more energetic inner parts \citep{kotov+2001,arevalo+2006}.

\citet{propagating_lag_paper} explore X-ray reverberation in self-consistent models that include such propagation through spatially extended coron\ae\ and show how the observed lag-energy profiles from X-ray reverberation reveal structure within the corona. In particular, they discuss how the characteristic dip seen in the lag-energy spectrum, with the earliest response seen in photons at 3\keV, rather than between 1 and 2\keV\ as would be na\"ively expected, is most likely explained by the upward propagation of high-frequency fluctuations through a collimated core of the corona, perhaps associated with the base of a jet, embedded within a more slowly-varying extended corona associated with the surface of the disc that is responsible for the low frequency propagation lags.

Some of the greatest insight into the accretion process has come from observations of a particular class of AGN, the narrow line Seyfert 1  (NLS1) galaxies. NLS1 galaxies are characterised by Doppler broadening of only around 2000\kmps\ in lines emitted from the broad line region and often show strong Fe\,\textsc{II} emission in their optical spectra but weak [O\,\textsc{III}] emission \citep{osterbroke_pogge}. There are strong indicators that NLS1 galaxies have relatively high accretion rates compared to typical Seyfert galaxies, but possess lower mass black holes than typical broad line Seyfert 1 systems \citep[\textit{e.g.}][]{boller+96}. A sub-class of the NLS1 galaxies, the complex NLS1s, those found to be relatively weaker in their X-ray emission compared to the optical/UV, exhibit extreme variability in their X-ray emission with the light curves showing peaks and sharp drops as well as spectral complexity above 2.5\keV, characteristic of reflection from the accretion disc or of absorption \citep{gallo_nls1}.

I Zwicky 1 (I\,Zw\,1, $z=0.0611$) is often considered the prototypical NLS1 galaxy from its optical properties. Observations of I\,Zw\,1 with \textit{XMM-Newton} in 2002 and 2005 showed a sharp drop in the X-ray count rate with evidence for two distinct modes of X-ray variability exhibited over different parts of the observation. The shape and normalisation of the spectrum were seen to vary before the drop in X-ray flux, but only the variability in the normalisation was manifested after. These modes suggested that two X-ray emission components coexisted in I\,Zw\,1; a diffuse component responsible for the long-term variability over the course of years and a compact component, perhaps associated with the base of a jet, responsible for the more rapid variability \citep{gallo_1zw1_1,gallo_1zw1_2}.

Following the discovery of this fascinating behaviour in I\,Zw\,1, we analyse in this work X-ray observations made by \textit{XMM-Newton} in 2015. In particular, we employ X-ray timing techniques to study the variability of the X-ray emission and structure and evolution of the corona to further understand these distinct coronal emission components. The emission from I\,Zw\,1 exhibits a flare-like increase in flux during the second orbit during which the X-ray count rate increased by around 30 per cent for a period of 50\ks.


\section{Observations}
I\,Zw\,1 was observed by \textit{XMM-Newton} \citep{xmm_strueder} during two orbits on 2015 January 21 and 22 (OBSIDs 0743050301 and 0743050801) with exposure totalling 275\ks. In this work, we use the light curves and spectra recorded by the EPIC cameras on board \textit{XMM-Newton}, predominantly the pn camera due to its enhanced effective area facilitating detailed X-ray timing analysis, but also the MOS cameras. MOS and pn data were analysed separately meaning we need not be concerned by the cross-calibration of the instruments. The observation was optimised for high resolution spectroscopy of outflows from the AGN using the \textit{Reflection Grating Spectrometer} (RGS), hence the source was moved to different locations around the centre of the field of view such that during each pointing, the bad pixels in the RGS camera fall at different energies. This enables measurement of the spectrum over the full RGS bandpass. This resulted in the pn observations being split into five segments during each orbit with an approximately 2.8\ks\ gap between each. The MOS cameras operated continuously, however, during each orbit. Detailed analysis of the RGS spectra will be presented in a future paper. 

Data were reduced using the \textit{XMM-Newton} \textsc{Science Analysis System} (\textsc{sas}) v15.0.0 using the most recent calibration data for the observations in question available at the time of writing. The event lists for each pointing were reduced following the standard procedure and time intervals during which the background count rate flared were removed using the standard criterion (that the total count rate between PI channels 10000 and 12000 in the pn detector is above 0.4\ctsps). Background flaring was particularly pronounced in the pn camera (though the MOS cameras were relatively unaffected) for periods during the second orbit with the filter reducing the usable exposure to 115.6\ks\ out of the 127.6\ks\ (or 91 per cent of the) total exposure time during the pointings of this orbit. The first orbit was not affected with 99.9 per cent of the exposure time usable.

We defined a circular source extraction region on the detector, centered on the point source, 35\,arcsec in diameter. A corresponding background region of the same size was selected on the same chip as the source. Light curves were extracted from the event lists. The background was subtracted and the count rate in each time bin corrected for exposure and dead time using the \textsc{sas} task \textsc{epiclccorr}. The spectra taken from the event lists were binned using the \textsc{grppha} tool such that there were at least 25 counts in each spectral bin and that the errors are approximately Gaussian. The photon redistribution matrices (RMF) and ancillary response matrices (ARF), encoding the effective area as a function of energy, were computed using the \textsc{sas} tasks \textsc{rmfgen} and \textsc{arfgen}, following the standard procedure.

\begin{figure*}
\centering
\includegraphics[width=170mm]{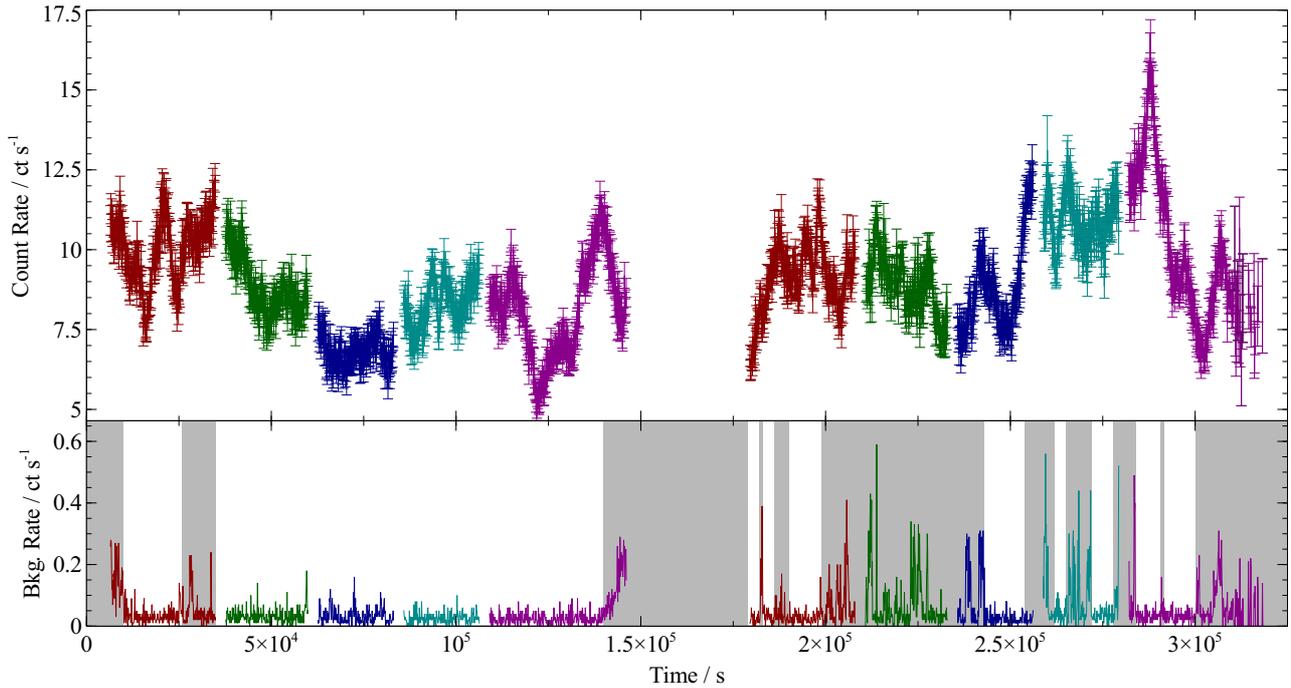}
\caption[]{EPIC pn light curve over the full 0.2 to 10\keV\ energy band of I\,Zw\,1 during the 2015 \textit{XMM-Newton} observations. Colours show the separate pointings with the first five taken during the first orbit and the last five during the second orbit. The lower panel shows the pn background count rate and the grey shaded regions show the time intervals excluded from the final timing analysis avoiding the periods of high background count rate.}
\label{lc.fig}
\end{figure*}

The total, exposure-corrected, light curve in the 0.2-10\keV\ \textit{XMM-Newton} bandpass is shown in Fig.~\ref{lc.fig}, with the separation between the pointings during each of the orbits shown. The average count rate detected from the source was 7.9\ctsps with a total of $1.4\times 10^6$ counts in the EPIC pn camera over the course of the observations. In each of the MOS1 and MOS2 cameras, the average count rate was 1.66\ctsps\ with a total of $1.1\times 10^6$ counts detected across both of the detectors.

\section{X-ray Timing Analysis}
In order to determine if the X-ray continuum of I\,Zw\,1 reverberates off of the inner regions of the accretion disc and to determine the causal relationship between emission components that make up the X-ray spectrum, the \textit{lag-frequency spectrum} was computed between the 0.3-1.0\keV\ energy band, where the soft excess composed of blurred emission lines originating from the ionised accretion disc is expected to be the most dominant and the 1.0-4.0\keV\ energy band, expected to be dominated by continuum emission that is observed directly from the corona. The lag-frequency spectrum shows the time lag between correlated variability in these two light curves for the different Fourier frequency components (\textit{i.e.} the slower and faster components that make up the total variability in the emission) that make up the light curves.

The power of X-ray timing analysis is that correlated variability is pulled out of the light curve, which means that reverberation from the accretion disc can often be picked out even when it is not obvious in the time-averaged X-ray spectrum, for example due to the presence of absorption or other emission components. Due to the form of the reflection spectrum, it will always contribute more to the observed emission between 0.3 and 1.0\keV\ than between 1.0 and 4.0\keV. Hence, even if it is not the dominant component in the former energy band, the fact that this band has a greater fractional contribution of reflected emission than the latter means that, on average, variability in the former will lag behind that in the latter, even though the time lag will be diluted from its intrinsic value were the bands to contain purely reflected and continuum emission respectively \citep{lag_spectra_paper,cackett_ngc4151}. Measuring the time lag as a function of Fourier frequency allows us to pick out the relationship between processes that induce variability on different timesales.

Light curves in these energy bands were extracted, then the cross-spectra were computed following the method outlined in \citet{reverb_review}, from which time lags were measured.\footnote{X-ray timing analyses were conducted using pyLag, available at \url{http://github.com/wilkinsdr/pylag}} Because the pn observation was taken as a series of individual pointing segments with short gaps in between, the cross spectrum was computed for each of the pointing segments separately and the average cross spectrum was computed for each frequency bin across all of the pointings. The 30\,ks duration of each pointing limits the lowest Fourier frequencies that can be probed using these continuous light curves.

The lag-frequency spectrum in these energy bands for I\,Zw\,1 is shown in Fig.~\ref{lagfreq.fig}. By convention, a positive time lag indicates that variability in the harder X-ray band lags behind that in the softer band, hence reverberation off the disc is indicated by negative time lags, where the softer reflection-dominated band is lagging behind the harder continuum-dominated band.

\subsection{The effect of background flaring}
Observations during the second \textit{XMM-Newton} orbit (OBSID: 0743050801) with the EPIC pn camera were, in part, affected by flaring in the background which can affect X-ray timing measurements in a number of ways, depending on the contribution of the flaring to the source light curves and the Fourier transforms thereof. Typically, during X-ray spectral analysis, periods of high background count rate are excluded from the accumulated spectrum. The same approach, however, when applied to spectral timing measurements results in gaps in the light curve, which depending on their length and frequency, will manifest across a number of Fourier components.

During the initial data reduction, the standard filtering criterion was applied to the EPIC pn event list, leaving a number of gaps scattered throughout the light curves obtained during the second orbit. After applying this event selection criterion, there were, however, still a number of short periods during which the background count rate was slightly above the relatively constant level that is expected.

Background flaring is not \textit{a priori} expected to induce additional lags between light curves since, presumably, when a flare hits the detector, the count rate is increased across all X-ray energies (or at least all affected energies) simultaneously. This does mean, however, that a component is added to the variability with zero-lag between affected light curves. While background flaring will not induce additional lags between spectral components, the addition of simultaneous variability between the bands act to dilute (\textit{i.e.} shorten) the measured lag over frequency ranges that are determined by not only the variability frequency of the emission during the flare itself, but the length of the flaring intervals and the gaps between them.

\begin{figure}
\centering
\includegraphics[width=80mm]{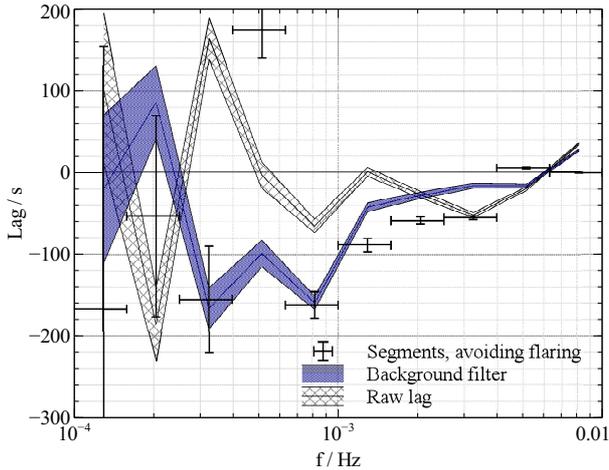}
\caption[]{Comparison of the lag-frequency spectrum when different techniques are used to mitigate background flaring. Initially, the raw lag was computed, treating each pointing as a single, continuous light curve (grey, hatched shaded region). Time periods of high background count rate were then excluded but leaving gaps in the light curves which were extrapolated over (blue, solid region). Finally, the lag spectrum was computed over 13 individual segments using only data from time intervals between the background flares (points).}
\label{lagfreq.fig}
\end{figure}

In order to understand the effect of background flaring on these observations, the lag-frequency spectrum was computed employing a number of methods to mitigate the flaring. Initially, the raw lag-frequency spectrum was computed in which the cross spectrum was simply computed between the light curves during each individual pointing and then averaged across the pointings. The raw lag-frequency spectrum, shown as the grey hatched region in Fig.~\ref{lagfreq.fig}, shows a double-dip structure at mid frequencies in which a negative lag, indicating that the soft band lags behind the hard, is seen initially from $5\times 10^{-4}$\Hz\ to $1.2\times 10^{-3}$\Hz. The lag then reduces to zero before turning negative again between $1.2\times 10^{-3}$\Hz\ and $6\times 10^{-3}$\Hz. The measured lag between the hard and soft bands is about 60\s\ over both of these frequency ranges.

As well as obtaining the raw lag-frequency spectrum, the lag-frequency spectrum was also computed after employing a stricter background filtering criterion. Over the course of the observations, except during flaring periods, the count rate from the background extraction region was seen to be consistently around 0.05\ctsps, thus sections were removed from the light curves when the background count rate increased above this level. The resultant gaps in the light curve were filled by linear interpolation from the last point before the gap to the first point after which, as discussed by \citet{reverb_review} does not impact the measured lag spectrum so long as the gaps are short in comparison to the full light curve and in comparison to the timescale on which the lag is being measured.

From Fig.~\ref{lagfreq.fig}, comparing the lag-frequency spectrum computed with the flaring intervals removed using the stricter criterion (the solid blue shaded region) to the raw lag-frequency spectrum (the grey hatched region), it can be seen that the double-dip structure and the zero-lag point at $1.2\times 10^{-3}$\Hz\ is an artefact of the background flaring inducing a zero-lagging component between the two light curves. Moreover, the measured lag below this point has increased to 160\s, indicating that below $1.2\times 10^{-3}$\Hz, simultaneous variability within the flaring periods was substantially diluting the measured lag.

In order to obtain a more robust measure of the lag spectrum, shorter light curves were extracted just from the sections of the observations between background flares. This is necessary since the longest gap left by filtering based on the background count rate was 1.7\ks\ and it cannot be guaranteed that such a long gap will not impact the measured lag spectrum in the region of interest where evidence of reverberation from the accretion disc is seen. The cross spectrum was computed for each of the light curve segments extracted from the time periods between the flares, then the lag was computed from the average cross spectrum across the segments. Because the light curve segments are now of different lengths, they cannot all contribute to the low frequency bins. In each frequency bin (central frequency $f$) in the lag-frequency spectrum, the average cross spectrum is taken between all light curves that are longer than $1/2f$ such that all segments contribute to the high frequency bins and only the longer ones contribute to the low frequency bins. Since the spacing of sample frequencies obtained from the discrete Fourier transform is inversely proportional to the length of the light curve segment, the contribution of each light curve segment to the average cross spectrum in each frequency bin is implicitly weighted by the length of the light curve segments when the average cross spectrum in each bin is calculated from all of the individual sample frequencies across all the segments that fall into that bin. The combined lag spectrum thus reflects the contribution from each time period had the cross spectrum been computed from the original, continuous light curve (the integral performed during a Fourier transform implicitly averages over the time series). A total of 13 light curve segments were used, ranging between 20 and 37\ks\ in duration with a total exposure during the second orbit of 51.1\ks\ out of the 127.6\ks\ original exposure.

The final lag-frequency spectrum, shown as points in Fig.~\ref{lagfreq.fig}, is largely consistent with that computed where the light curves were filtered based on the background count rate (leaving gaps) below (above $10^{-3}$\Hz). Leaving the gaps in the filtered light curve, however, causes the higher frequency lags (above $1.2\times 10^{-3}$\Hz) to be diluted from 60\s\ to 20\s. Over the frequency range $3\times 10^{-4}$\Hz\ to $6\times 10^{-3}$\Hz, variability in the soft, 0.3-1\keV\ band is seen to lag behind that in the hard, 1-4\keV\ band. Over the lower part of this frequency range, the time lag is $(160\pm 16)$\s, but at $1.2\times 10^{-3}$\Hz, the lag time drops to $(59\pm 4)$\s\ (with the lags measured in the frequency bins at $8\times 10^{-4}$ and $2\times 10^{-3}$\Hz\ respectively). The soft lag is detected over a range in frequencies in these observations of I\,Zw\,1 with a transition between two distinct lag times. At low frequencies, below $3\times 10^4$\Hz, there is evidence of an up-turn in the lag-frequency spectrum, with the hard band lagging the soft, however, due to the relatively short light curve segments, it is not possible to fully probe these low frequencies.

We note that there is an anomalous point in this final lag-frequency spectrum around $5\times 10^{-4}$\Hz\ where a single frequency bin shows a sharp transition to a positive lag from the surrounding negative values. Further investigation of the lag-frequency spectrum employing different frequency binning reveals that this anomaly is restricted to a very narrow range of frequencies. This frequency corresponds to a timescale around 2\ks\ which is of order the length of the gaps between the pn pointing segments. This should not obviously affect the lag measurements since we do not interpolate over the gaps, rather stack the lag spectra from the individual segments. We surmise it is likely due to residual noise, correlated between the energy bands, after the flaring was removed influencing this timescale, perhaps due some slight variability between the segments. Henceforth, this frequency bin is excluded from our analysis.

\begin{figure}
\centering
\includegraphics[width=80mm]{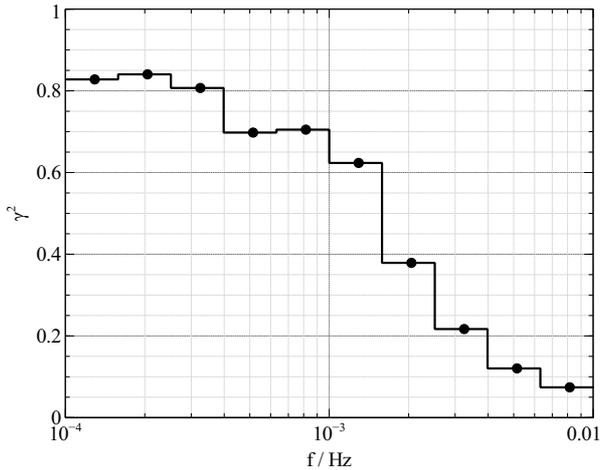}
\caption[]{The raw coherence between light curves over the 0.3-1.0\keV\ and 1.0-4.0\keV\ energy bands used in the computation of errors on the lag measurements between these light curves.}
\label{coherence.fig}
\end{figure}

Errors on the time lags are calculated from the coherence function, $\gamma^2$. The coherence between two light curves represents the fraction of the variability from one light curve that can be predicted by a linear transformation from the other (\textit{i.e.} the application of the response function that describes the time delays in the reflected emission relative to the continuum). $\gamma^2=1$ represents perfect coherence between the two time series and as more uncorrelated variability is introduced (\textit{e.g.} Poisson noise that affects each of the light curves independently), the coherence is reduced. The coherence between the 0.3-1\keV\ and 1-4\keV\ light curves for which the lag-frequency spectrum was computed is shown in Fig.~\ref{coherence.fig}. A drop in coherence is seen above $1.5\times 10^{-3}$\Hz. \citet{epit+16} show that where the cross spectrum is computed over $m$ light curve segments, systematic errors on lag measurements are minimal (\textit{i.e.} the intrinsic lag between the light curves and the corresponding error bar) is accurately represented so long as $\gamma^2 > 1.2 / (1 + 0.2m)$. For the 13 light curves used here in producing the final lag-frequency spectrum, the coherence should be greater than 0.3 for a reliable measure of the lag, hence the two reverberation lag times measured below $(2\sim 3)\times 10^{-3}$\Hz\ are reliable. Below this frequency, as the contribution of Poisson noise to the power spectrum increases, the effect is only to enlarge the error bars on the lag measurement, without skewing the measured value.

Although the EPIC MOS light curves were not significantly affected by background flaring during these observations, we find that there is insufficient signal-to-noise due to the smaller effective area of these detectors to produce robust measurements of time lags between energy bands.

\subsection{Signatures of reverberation}
The lag-frequency spectrum of I\,Zw\,1 exhibits the features typical of Seyfert galaxies. At the lowest frequencies, despite the large error bars, there is evidence of an up-turn towards positive lags. This suggests that in these slowest components of the variability, the hard band lags behind the soft. This behaviour is observed across AGN and was first discovered in X-ray binaries. Rather than being due to reverberation from the accretion disc, this behaviour is observed in sources that do not display significant X-ray reflection \citep[\textit{e.g.}][]{walton_hardlag} and is posited to arise within the coronal continuum emission itself. In many Seyfert galaxies, the time lag at the lowest frequencies is seen to rise steadily with increasing energy bands \citep{kara+13} as the variability in coronal emission responds first at lowest energies then later at higher energies, a behaviour that can be explained by the coronal luminosity being modulated by fluctuations in the mass accretion rate propagating inwards through the accretion disc \citep{kotov+2001,arevalo+2006}. The less-energetic outer parts of the corona respond initially but the more energetic inner parts that dominate the harder continuum emission do not respond until later as it takes time for the fluctuation to propagate in through the disc. Due to the short pointings from which continuous light curves could be extracted, it is not possible to probe the low frequency behaviour of the lag spectrum in detail as a function of energy.

Negative time delays indicative of reverberation from the accretion disc (where the hard, continuum band \textit{leads} the soft reflection-dominated band) are seen at higher frequencies. Typically, the lag-frequency spectrum of Seyfert galaxies shows a single dip or range of frequencies over which reverberation from the disc dominates the time lags over which it smoothly rises either side of the longest lag time. In these observations, however, two flattened regions of the lag-frequency spectrum are evident, with the lag dropping from 160\s\ over lower frequencies at $1.2\times 10^{-3}$\Hz\ to 60\s. Preforming rough fits to the lag-frequency spectrum from $3\times 10^{-4}$ to $4\times 10^{-3}$\Hz\ with a variety of simple phenomenological models show that due to the position of the data point at $8\times 10^{-4}$\Hz\ a step function in lag as a function of frequency is preferred over a smooth decrease in lag towards higher frequencies described by either linear, log-linear or power law functions ($\Delta\chi^2 = 10$ but with only three degrees of freedom, so it is difficult to state the preference of a step function with any great certainty).

In order to investigate this behaviour, lag-energy spectra were extracted, averaging the cross spectrum (and hence time time lag) between light curves in narrow energy bands and a reference band over the frequency range in question. In order to maximise signal-to-noise, the reference band for each energy bin was taken to be the full 0.3-10\keV\ range but with the present energy bin of interest excluded so as not to pick up correlated errors between the two bands. Lag-energy spectra were extracted using the same light curve segments as were used in constructing the final lag-frequency spectrum, avoiding periods of background flaring. In the case of the lag-energy spectra, the time lag is relative to the reference band, hence the relevant quantity is the lag of each energy band with respect to the others.

\begin{figure}
\centering
\subfigure[Lag-frequency Spectrum] {
\includegraphics[width=80mm]{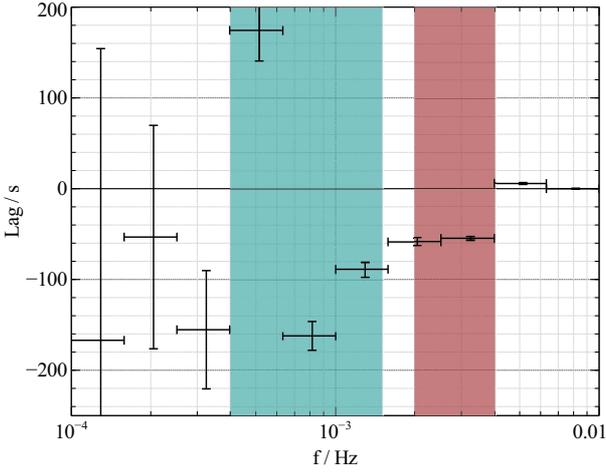}
\label{lagspec.fig:lagfreq}
}
\subfigure[Lag-energy spectrum, $0.4 < (f / 10^{-3} \mathrm{Hz}) < 1.5$] {
\includegraphics[width=80mm]{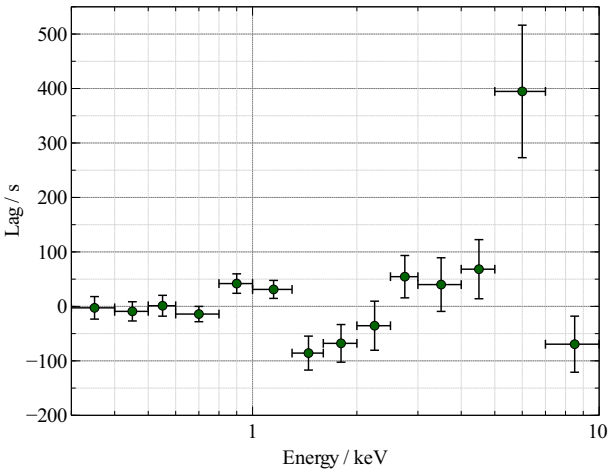}
\label{lagspec.fig:lagen_low}
}
\subfigure[Lag-energy spectrum, $2 < f / (10^{-3} \mathrm{Hz}) < 4$] {
\includegraphics[width=80mm]{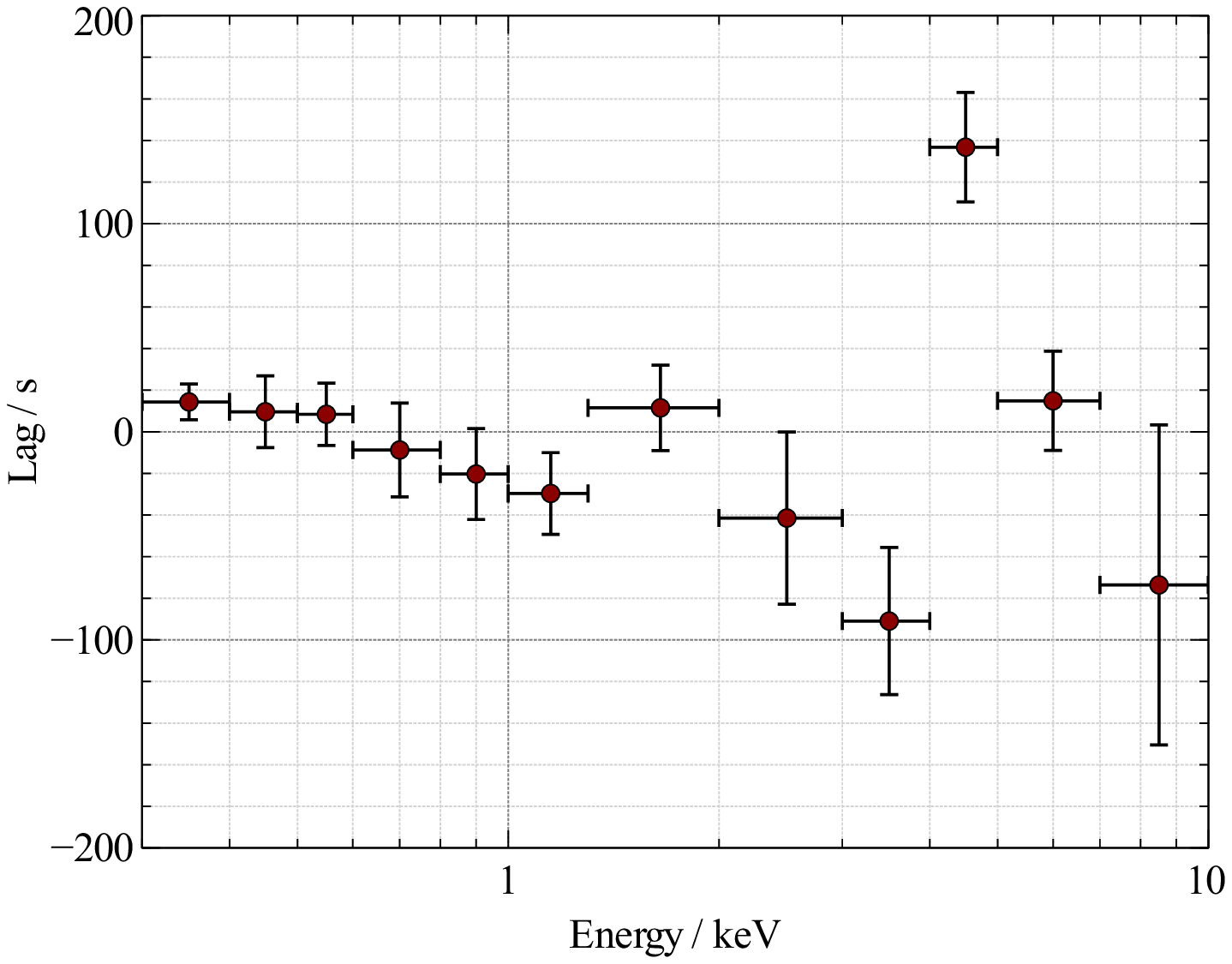}
\label{lagspec.fig:lagen_high}
}
\caption[]{\subref{lagspec.fig:lagfreq} The final lag-frequency spectrum from the 2015 \textit{XMM-Newton} observations of I\,Zw\,1 from light curves extracted over time periods avoiding flaring in the background with the frequency ranges exhibiting the longer and shorter reverberation lags (where the soft band lags behind the hard) highlighted. \subref{lagspec.fig:lagen_low} The lag-energy spectrum over the mid-frequency range over which the longer reverberation lag is seen and \subref{lagspec.fig:lagen_high} over the high-frequency range exhibiting the shorter reverberation lag.}
\label{lagspec.fig}
\end{figure}

The lag-energy spectra over the frequency ranges of the longer and shorter soft lags are shown in Figs.~\ref{lagspec.fig:lagen_low} and \subref{lagspec.fig:lagen_high} respectively. Due to additional noise resulting in a drop in coherence between 1.5 and 3\keV, it was necessary to rebin the light curves in energy in order to increase the signal-to-noise in the high frequency lag-energy spectrum. The lag-energy spectrum shows the relative response time of each energy band to variability that is correlated between the different bands. The zero point is arbitrary and depends upon the choice of reference band.

Both of these lag-energy spectra show the characteristic features of X-ray reverberation from the inner accretion disc. The lag-energy spectrum over the higher frequency range, shown in Fig.~\ref{lagspec.fig:lagen_high}, is typical of those seen due to reverberation in Seyfert galaxies. The response of the soft excess reflected from the disc below 1\keV\ as well as the iron K$\alpha$ fluorescence line around 6\keV\ (the rest frame energy of the line is 6.4\keV, but it is shifted and broadened by relativistic effects in the disc around the black hole) is seen to lag behind those energy bands dominated by directly observed continuum emission. The earliest response is seen in a suggestive dip at 3\keV, a feature commonly seen in lag-energy spectra due to reverberation from the inner accretion disc \citep{kara+13}.

Although commonly seen, this dip in the lag-energy spectrum at 3\keV\ is difficult to explain with simple models of reverberation from either compact point-like X-ray sources or coron\ae\ that extend over the surface of the accretion disc. The origin of this dip can however be understood in the context of reverberation of X-rays that originate from a collimated corona extending up the rotation axis of the black hole, akin to the base of a jet, through which fluctuations in luminosity propagate upwards, originating from either the inner regions of the accretion flow or the black hole magnetosphere \citep{propagating_lag_paper}. If luminosity fluctuations propagate up the extended corona sufficiently slowly ($\sim0.01c$), the 3\keV\ dip is produced since the X-rays that are emitted from the base of the vertically extended corona, close to the black hole, tend to be bent down onto the inner regions of the disc and hence are seen wholly in the reflected emission, not the directly observed continuum. The earliest emission in the iron K$\alpha$ line is seen at 3\keV\ since the photons that are redshifted down to this energy are reflected from the region of the disc with the shortest total light travel time (in the curved spacetime around the black hole) from the source to the disc to the observer. More extremely redshifted photons originate from smaller radii but their passage is delayed in the strong gravitational field close to the black hole, whereas photons closer to the rest-frame energy of the line are reflected further out in the disc, hence the light travel time from the corona is longer. While usually the earliest response is seen in the continuum-dominated 1-2\keV\ band, sufficiently slow upward propagation means there is a delay before the fluctuation reaches a part of the corona from which a significant number of photons can escape to be observed in the continuum rather than being bent down towards the disc. Hence, in this scenario, the earliest average response is seen in the 3\keV\ energy band dominated by the prompt reflection from the disc. Alternatively, \citet{chainakun+2016} show that the 3\keV\ dip can arise from a point source a significant height above the black hole ($\gtrsim 5$\rg) reverberating from a disc with a steep ionisation gradient, with the ionisation parameter rising to $\xi = 10^4$\ergcmps\ at the inner edge.

On the other hand, the lag-energy spectrum over the frequency range of the longer `reverberation' lag, from $3\times 10^{-4}$\Hz\ to $1.2\times 10^{-3}$\Hz, clearly shows the profile that is na\"ively expected from reverberation from either a point-like corona or one extended over the surface of the disc through which luminosity fluctuations propagate inwards on the viscous timescale in the underlying disc (evaluating the model over the appropriate frequency range), as shown by \citet{propagating_lag_paper}. The earliest response is seen in the 1-2\keV\ band, dominated by emission directly observed from the corona. A delayed response is seen in the soft excess, below 1\keV, and in the iron K$\alpha$ line (relative to the earliest arriving photons at 1.5\keV), due to the additional light travel time between the corona and the reflecting disc. The redshifted wing of the iron K$\alpha$ line is seen to respond earlier than the core of the line \citep[see also][]{zoghbi+2013,cackett_ngc4151} since these redshifted photons originate predominantly from the inner regions of the disc, closer to the X-ray source (or the more rapidly varying inner parts of a corona extended over the disc) whereas the core of the line arises predominantly from the outer regions of the disc, further from the site of coronal emission. The detection of this lag-energy profile at low frequencies alongside the profile that is seen higher frequencies suggests the detection of a vertically collimated structure embedded within an extended structure associated with the accretion disc rather than the alternative explanation of a high point source and a steep disc ionisation gradient since both of the lag-energy profiles detected must be originating from the inner regions of the same accretion disc, hence a steep ionisation gradient cannot be invoked to produce one profile but not the other.

Note that in order to infer the geometry of the corona, following the findings of \citet{propagating_lag_paper}, after confirmation that the lag-energy spectrum exhibits the form expected for reverberation from the accretion disc (with a delayed soft excess and iron K$\alpha$ line with respect to the continuum), we focus on the location of the dip in the lag-energy spectrum between 1.5 and 3\keV\ which has been shown to provide the required information about the corona by itself. While the $2\sim 10$\keV\ lag-energy spectrum is found to be consistent between Seyfert galaxies (with this energy band being dominated by the continuum and its reflection from the disc), the spectrum can be more complex and divergent between different AGN below 1\keV\ \citep{kara+13}. The precise shape of the lag-energy spectrum below 1\keV\ will depend on the contribution of all of the emission components that contribute to this band in addition to the soft excess and the relative time lags between them.

\subsection{Detection significance}

In order to assess the significance of the detection of the dips in the lag-energy spectra, the lag as a function of energy over each of the two frequency bands was fit using a simple phenomenological model that was built up from a log-linear description of the underlying `continuum' of lags; the smooth variation in time lag as a function of energy underlying any spectral features (linear and power law descriptions were also tried, but these make no difference to the results). We find it is necessary, first of all, to include a Gaussian profile centred around 6\keV\ to account for lag in the centroid of the iron K$\alpha$ line, though formally if the lag continuum is allowed to vary freely (and pass through the 6\keV\ point), this peak is only mildly significant. The dip is fit by adding another Gaussian component to the lag-energy spectrum with free normalisation, centroid, and width. The best fitting centroid energy of the dip is $(1.62^{+0.03}_{-0.14})$\keV. In the case of the high-frequency reverberation signal, the dip is less significantly distinguished from the smoothly-varying lag `continuum.' The best fitting centroid energy of the dip is $(3.7^{+1}_{-0.4})$\keV\ with errors at the 90 per cent confidence limit.

To more robustly assess the significance of the detection of the dips in the lag-energy spectrum (\textit{i.e.} finding the energy band that responds earliest to the variations in count rate, the signature that is related to the geometry of the component of the corona in which the variability arises at the given frequencies), we performed Monte Carlo re-sampling of the observed light curves in each energy band. For each light curve, the photon count in each time bin was drawn from a Poisson distribution with mean equal to the measured photon count at that time. The light curves were re-sampled 50,000 times and the lag-energy spectrum was computed from each set of light curves. In order to determine the energy at which the reverberation response is the earliest in a manner independent of any model for the shape of the dip, the energy band, below 6\keV\ (since the 8-10\keV\ energy band is seen, almost ubiquitously among Seyfert galaxies, to respond first) with the lowest lag time was found for each of the re-sampled lag-energy spectra. In the case of the low frequency lag-energy spectrum, we find that in 92.4 per cent of the sample, the earliest responding energy bin lay between 1 and 2\keV. Comparing the average response time of the 1-2\keV\ and 2-4\keV\ energy bands, we find that in 99.7 per cent of the sample, the 1-2\keV\ band responds sooner, indicating that this dip is detected significantly above any random variation in the lag-energy spectrum due to random noise in the light curves.

The dip at 3\keV\ in the high frequency lag-energy spectrum is less significantly detected. In 83.9 per cent of the sample, the earliest overall response (below 6\keV) was seen between 2 and 4\keV (66.8 per cent have the earliest response in the more limited 3-4\keV\ range). In 93.1 per cent of the sample, energy bins between 2.5 and 4.5\keV\ in the high frequency lag-energy spectrum responded earlier than those between 1 and 2\keV. This lag-energy profile, however, is of the form expected from other observations of Seyfert galaxies where the dip has commonly been seen at 3\keV, thus the key result in I\,Zw\,1 is that the low frequency lag-energy spectrum significantly deviates from this form at the $3\sigma$ level and that the low and high frequency lag-energy spectra differ at the 93.1 per cent confidence level.


\section{Principal Component Analysis}
In order to gain a deeper understanding of the mechanisms of variability in the X-ray spectrum of I\,Zw\,1, we perform principal component analysis (PCA). PCA describes the variability of a data set (in this case the variation in the number of photon counts in different spectral channels over different time bins) as efficiently as possible in terms of a set of eigenvectors or principal components (PCs), such that the spectrum in any given time bin is expressed as the linear combination of the principal components each with a time-varying coefficient. Following the method of \citet{parker_pca}, PCA is performed using a singular value decomposition (SVD) algorithm to factorise a matrix that is constructed from the spectrum in successive time bins (with spectral channel along one axis and time bin along the other) and to obtain the eigenvectors.\footnote{PCA was performed using the code by Michael Parker, \url{http://www-xray.ast.cam.ac.uk/~mlparker}} The observations were divided into 1000\s\ time bins to perform PCA, though the results are consistent when time bins between 500 and 2000\s\ are used.

\begin{figure}
\centering
\includegraphics[width=80mm]{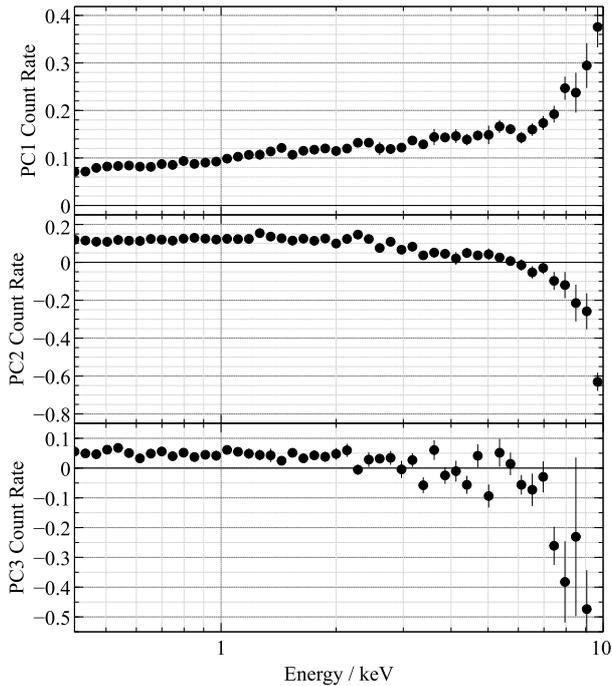}
\caption[]{The three significant eigenvectors (\textit{i.e.} those accounting for variability above the random noise) obtained from principal component analysis (PCA) of the X-ray spectral variability. PC1 accounts for 43 per cent of the total variability, PC2 13 per cent and PC3 6 per cent.}
\label{pca.fig}
\end{figure}

PCA is performed on the fractional variability from the mean in each spectral channel and the components obtained can be interpreted in the sense of energies that vary together in a correlated sense. In general, the SVD algorithm will produce as many principal components as there are time bins, however many of these will represent only uncorrelated random noise in each channel. The contribution of each component to the overall variability is represented by its corresponding eigenvalue. For random noise, the eigenvalues follow a geometric series, ordered from the highest (for the component contributing the most variability) to the lowest. Only those components with eigenvalues that are significantly greater than this geometric series given their error bars represent significant detections of underlying variability in the X-ray emission. Three such components were found in the case of I\,Zw\,1, the spectra of which are plotted in Fig.~\ref{pca.fig}, in descending order of their contribution to the total variability.

The first component, PC1, accounts for 43 per cent of the total variability. This component primarily accounts for the overall change in normalisation of the spectrum between the time bins. Since this component is positive across the entire energy range, the count rate in each spectral bin is increasing (or decreasing) together, though from the turn-up at high energies, it is clear that the fractional variability, on average, is greater in these channels. The PCA algorithm will attempt to describe as much of the variability as possible in the first component, hence it does not directly correspond to the variation in one spectral parameter, and application of the other components `corrects'  the spectrum to reproduce the observed shapes at different times.

\citet{parker_pca} show that PC2 (which represents 13 per cent of the total variability) is of the form observed when in addition to the normalisation varying, the slope of the continuum component (the photon index) varies. The change in sign above 6\keV\ shows that these higher energies are anti-correlated with the low energy portion of the spectrum, thus reproducing the pivoting of the spectrum as the photon index varies.

PC3, representing 6 per cent of the variability, shows the same shape as PC2, with anti-correlation between high and low energy portions of the spectrum, suggesting that this component also corresponds to variation in the slope of the continuum spectrum. Its appearance as a separate component (although being the same shape as PC2) suggests that while this variability produces approximately the same changes in the spectral shape, it is not correlated with that accounted for by PC2. The appearance of this second component suggests that two power law continuum components contribute to the observed spectrum but that their variability is not correlated on the 1000\s\ timescale on which the spectra were binned for PCA. The separation of PC2 and 3 support the interpretation of the X-ray timing analysis, that the corona is separated into distinct structures; the extended and inner parts driven by different stochastic variability.


\section{Evolution of the Corona}
Studying the lag-frequency spectra from each of the two orbits of \textit{XMM-Newton} individually in conjunction with the light curve (Fig.~\ref{lc.fig}) reveals evidence for evolution in the structure of the corona and the driving force behind the variability. Most notably, there is a flare in the X-ray emission apparent in the light curve between the fourth and fifth pointings of the second orbit. The count rate initially increased by 15 per cent from the mean level up until this point, then stabilised at this increased level for 29\ks\ before increasing a further 44 per cent during a sharp peak lasting 17\ks at the end of the flare, after which the X-ray count rate returned to its pre-flare level.

\begin{figure*}
\begin{minipage}{175mm}
\centering
\subfigure[First orbit] {
\includegraphics[width=80mm]{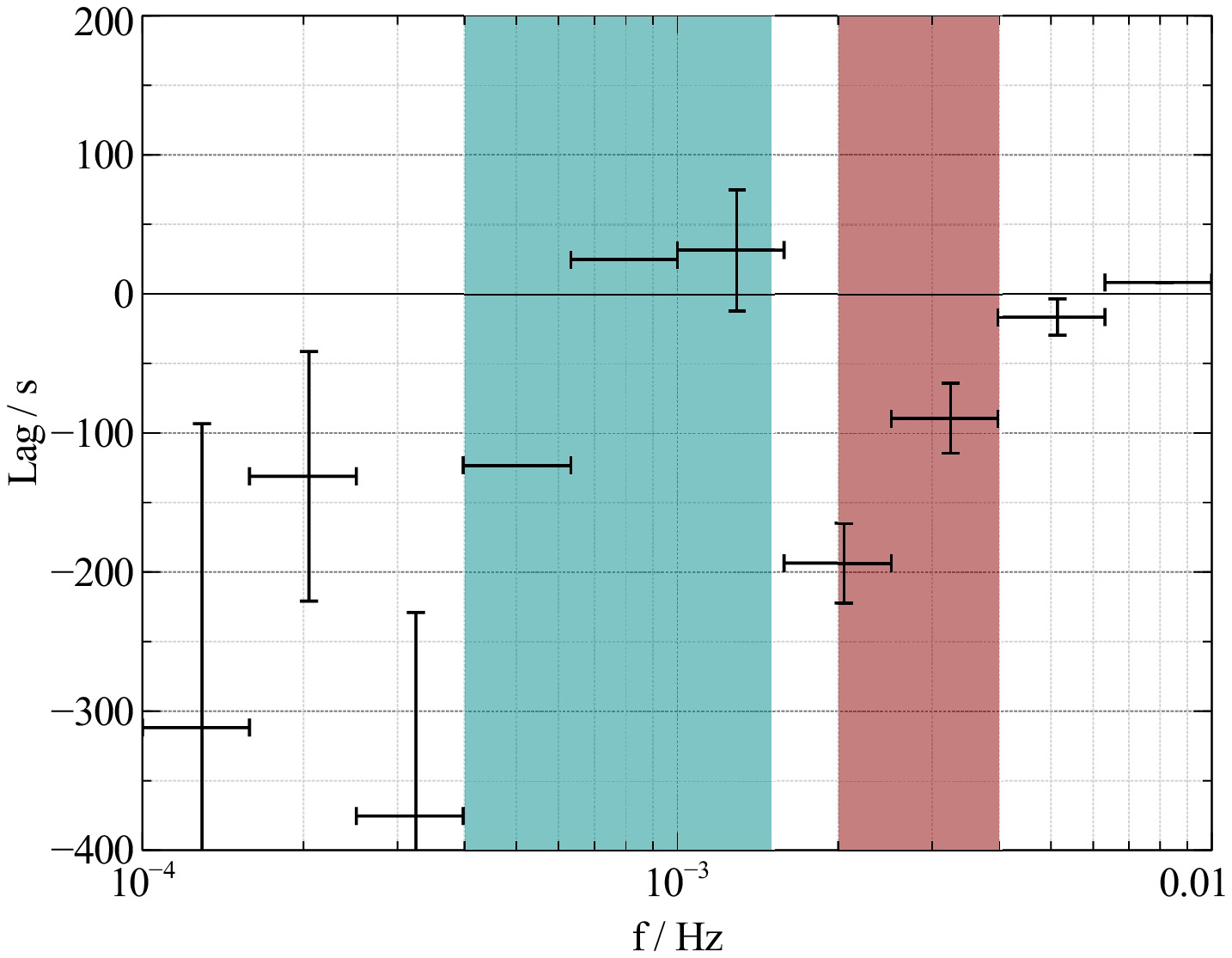}
\label{lagfreq_pointings.fig:orb1}
}
\subfigure[Second orbit] {
\includegraphics[width=80mm]{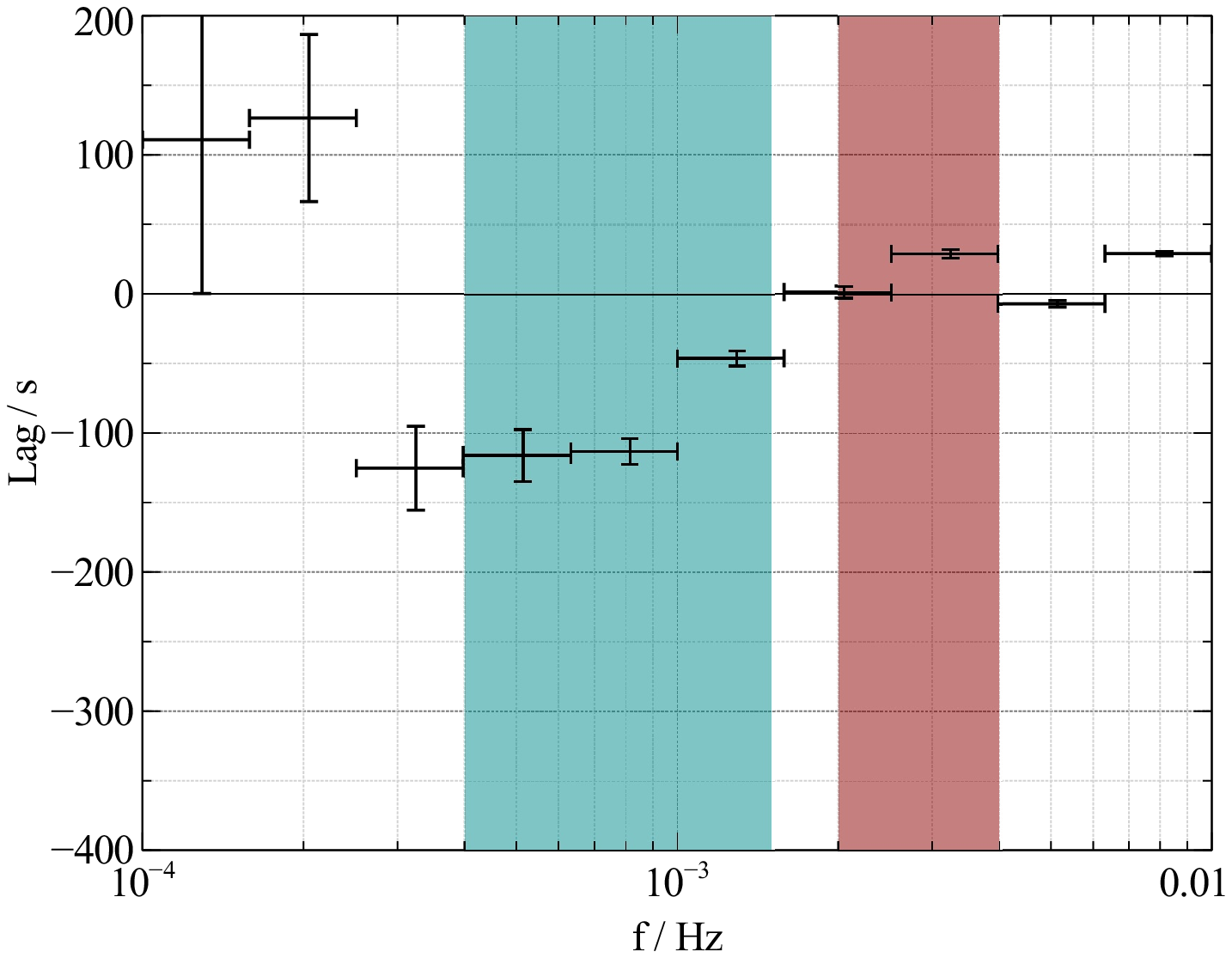}
\label{lagfreq_pointings.fig:orb2}
}
\caption[]{Comparison of the lag-frequency spectrum obtained from the first and second orbits. The frequency ranges over which the low and high frequency reverberation signals were detected previously from the combined lag spectrum are highlighted.}
\label{lagfreq_pointings.fig}
\end{minipage}
\end{figure*}

Fig.~\ref{lagfreq_pointings.fig} shows the lag-frequency spectrum over the first and second orbits (with the frequency ranges over which the two lag-energy profiles were seen highlighted for reference). While it was possible to extract lag-frequency spectra for each of the orbits individually, it was not possible to further sub-divide the observation for the purposes to maintain the required signal-to-noise for X-ray timing analysis, hence it was not possible to find if the spectra are stable within each orbit or directly study the evolution of the lags on shorter timescales.

\begin{figure}
\centering
\includegraphics[width=80mm]{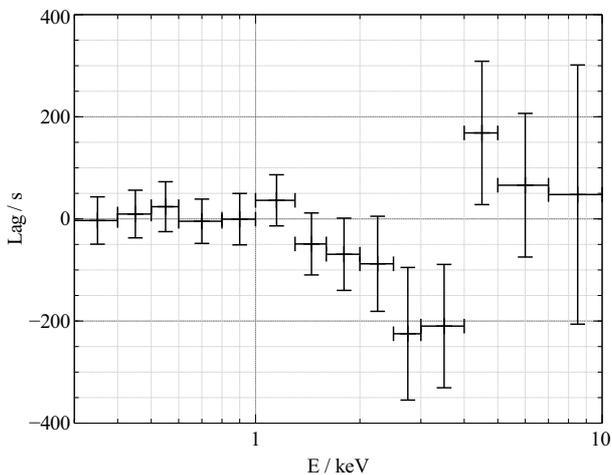}
\caption[]{The lag-energy spectrum extracted from the frequency range $(0.4-4)\times 10^{-3}$\Hz\ from only the first orbit of the observation, showing the dip at 3\keV\ indicative of upward propagation through a collimated core to the corona.}
\label{lagen_orbit1.fig}
\end{figure}

The X-ray reverberation lags, particularly those over the lower frequency range associated with a component of the corona that extends over the surface of the accretion disc, with the lag-energy spectrum showing a dip between 1 and 2\keV, are most clearly detected during the second orbit. The reverberation lag time of $\sim 120$\s\ over the $(3-15)\times 10^{-4}$\Hz\ band is clearly visible in Fig.~\ref{lagfreq_pointings.fig:orb2}, while it does not appear in Fig.~\ref{lagfreq_pointings.fig:orb1} for the first orbit. On the other hand, the higher frequency soft lag is seen most clearly during the first orbit. It was over this frequency band that the dip in the lag-energy spectrum was detected at 3\keV\ and the appearance of this dip during the first orbit, without evidence for the 1-2\keV, dip can be seen from the lag-energy spectrum from only the first orbit, over the entire $(4-40)\times 10^{-4}$\Hz\ frequency range (the range over which \textit{both} of the reverberation components were seen when data from the two orbits were combined), shown in Fig.~\ref{lagen_orbit1.fig}. This 3\keV\ dip is associated with a collimated core in the corona through which fluctuations in luminosity propagate upwards.

\subsection{Evolution of the power spectrum}
\begin{figure}
\centering
\includegraphics[width=80mm]{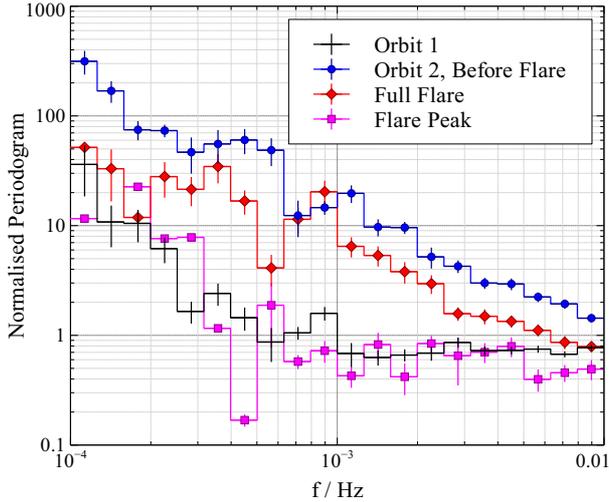}
\caption[]{The normalised periodogram, representing the power spectral density of variability in the X-ray light curve measured with the EPIC MOS camera between 0.3 and 10\keV. The periodogram is shown for segments of the observation; the first orbit, the period before the flare during the second orbit, the full flare and just the peak in emission at the end of the flare.}
\label{periodogram.fig}
\end{figure}

The appearance of the lower frequency reverberation signal, associated with the component of the corona extended over the disc, can be understood in terms of the power spectral density (PSD) of the variability as a function of Fourier frequency, represented by the periodogram, appropriately normalised \citep{reverb_review}. Since the pn observations are affected by background flaring, we use the light curves measured by the EPIC MOS cameras (summing the light curves from MOS1 and MOS2) to compute the periodogram. While the signal-to-noise in these detectors proved insufficient for the lag analysis, it is sufficient to measure the PSD. The MOS cameras not having been substantially affected by the background flaring during these observations means variability induced in the background will not inadvertently be attributed to variability in the source. The normalised periodogram is shown in Fig.~\ref{periodogram.fig} for segments of the observation; the first orbit, the period before the flare during the second orbit, the full flare and just the peak in emission at the end of the flare.

It should be noted that the periodogram that is measured from short light curve segments can be influenced by the effects of red noise leak where variability at frequencies lower than those sampled in the discrete Fourier transform of the short light curve segments is manifested at higher frequencies, at the bottom end of the range of frequencies that are sampled in the measured periodogram. Red noise leak becomes more apparent the more intrinsic variability there is below the sampled frequency range, thus the further the break frequency below which the PSD flattens (rather than continuing to rise) lies below the minimum sample frequency, the more variability power lies outside the sampled bandpass and the more significant the red noise leak. The extra variability that is inferred at low frequencies mean that red noise leak can cause the slope of the PSD to be over-estimated.

While it is difficult to mitigate the effects of red noise leak from the shortest light curve segment during the flare, in addition to measuring the periodogram from the full light curve over each of the intervals, we divide each of the intervals into segments equal to the length of the shortest interval; the peak of the flare (16\ks), the compute the periodogram averaged over the appropriate frequency points over all of the segments within each frequency bin. This ensures that, given the same intrinsic PSD, the effect of red noise leak is consistent between each of the periodograms we measure and thus that the variability we see in the periodogram over time is due to intrinsic changes in the PSD rather than the changing effects of red noise leak. We find that the periodograms computed from the segmented light curves in each time interval are consistent with those computed from the full light curves.

It is clear that the variability power falls off much more rapidly during the first orbit than in the second. During the first orbit, the periodogram falls off as approximately $f^{-2}$ over the range $f<5 \times 10^{-4}$\Hz\ (the best fitting power law index over this range is $2.1_{-0.6}^{+0.5}$), above which the variability power is constant across frequencies. Such a flat power spectrum is associated with the Poisson noise in the photon arrival rate. During the second orbit, the variability rises dramatically, even before the onset of the flare. The best fit power law to the periodogram during the period of the second orbit before the flare falls off as $f^{-1.1}$ over the full range of measured frequencies with a best fitting index of $1.08_{-0.03}^{+0.03}$ (although the peridogram reaches the Poisson noise level at the uppermost frequency that was probed). This enhanced variability continues as the flare begins with the best fitting power law to the periodogram falling off with index $0.89_{-0.05}^{+0.05}$.

The appearance during the second orbit of the low frequency reverberation signal associated with the extended portion of the corona can be understood in regard to the variability observed in the PSD. During the first orbit, there is significantly less power associated with variability over the low frequency range associated with reverberation from the extended portion of the corona. The extended portion of the corona is not necessarily absent during the first orbit, rather its emission is more variable during the second orbit, meaning the reverberation at low frequencies is more easily detected; there is a significant coherent signal between the soft and hard bands. During the first orbit, the corona is less variable across all frequencies. More powerful variability is required to obtain the equivalent signal-to-noise level at low frequencies than at high frequencies since over the course of a single observation, fewer wave cycles of the lower frequency components are available to average over in the cross spectrum. If a significant portion of the variability that is there is now at high frequencies and arising arising in the collimated core of the corona, we see the lag-energy profile due to this coronal structure. It is the lower variability power during the first orbit that is responsible for the drop in coherence between the light curves, leading to the larger errors seen on these lag spectra as well as the scatter at low frequencies (lower frequencies are less well sampled during the observation as fewer wave cycles of the variability are available over which lag measurements can be conducted). Lower photon count rates would also reduce the signal-to-noise of the variability and the time lags therein, but in this case we see that the reverberation signal from the extended corona becomes apparent in the second orbit even before the count rate increases during the flare, suggesting that it is available variability power that is the determining factor on whether the reverberation signal can be detected.

We note, however, that during the peak in X-ray count rate at the end of the flare, the variability has already dropped back to its pre-flare level, even though the count rate rises again before dropping sharply.

\subsection{Evolution of the variable spectral component}

Further insight into the evolution and structure of the corona can be gained from the \textit{covariance} spectrum, calculated as is outlined in \citet{reverb_review}. The covariance spectrum is analogous to the RMS or fractional variability spectrum, showing the relative levels of variability in successive energy bands, hence revealing the shape of the variable component in the X-ray spectrum. Rather than just measuring the variability in a single light curve for each energy band, however, the covariance spectrum is computed from the magnitude of the cross spectrum and measures the variability that is correlated with a reference band which removes the contribution from uncorrelated noise. The reference band essentially acts as a matched filter for the variability of interest. As for the lag-energy spectra, the reference band was taken to be the summed light curve over all energies minus the present band of interest so as to avoid correlated noise. As for the periodogram, there was sufficient signal-to-noise in the EPIC MOS data for the covariance spectra, so we use the summed MOS1 and MOS2 light curves to avoid the contribution of background flaring to the variability.

\begin{figure*}
\begin{minipage}{175mm}
\centering
\subfigure[$0.4 < (f / 10^{-3} \mathrm{Hz}) < 1.5$] {
\includegraphics[width=80mm]{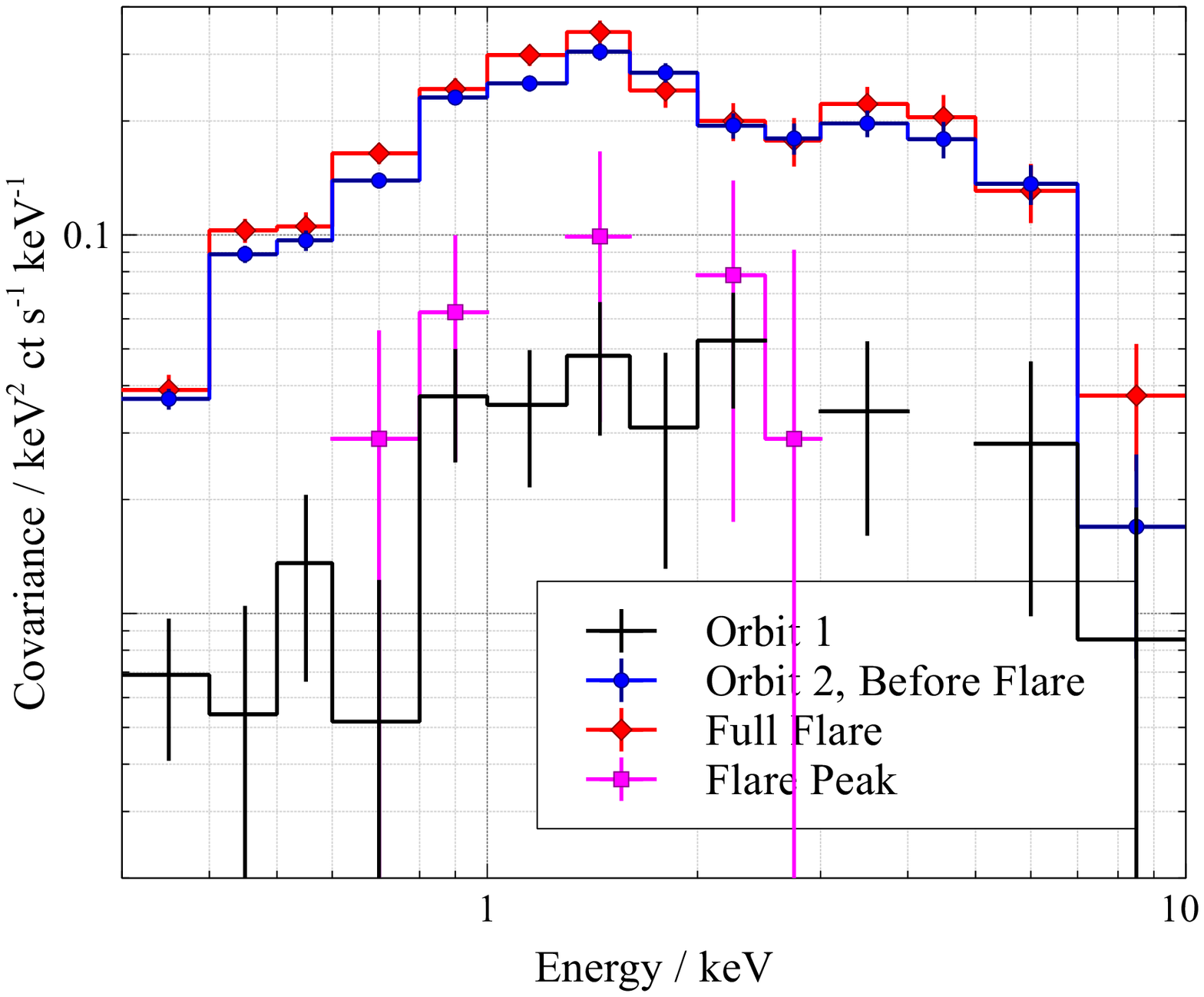}
\label{covariance.fig:low}
}
\subfigure[$2 < f / (10^{-3} \mathrm{Hz}) < 4$] {
\includegraphics[width=80mm]{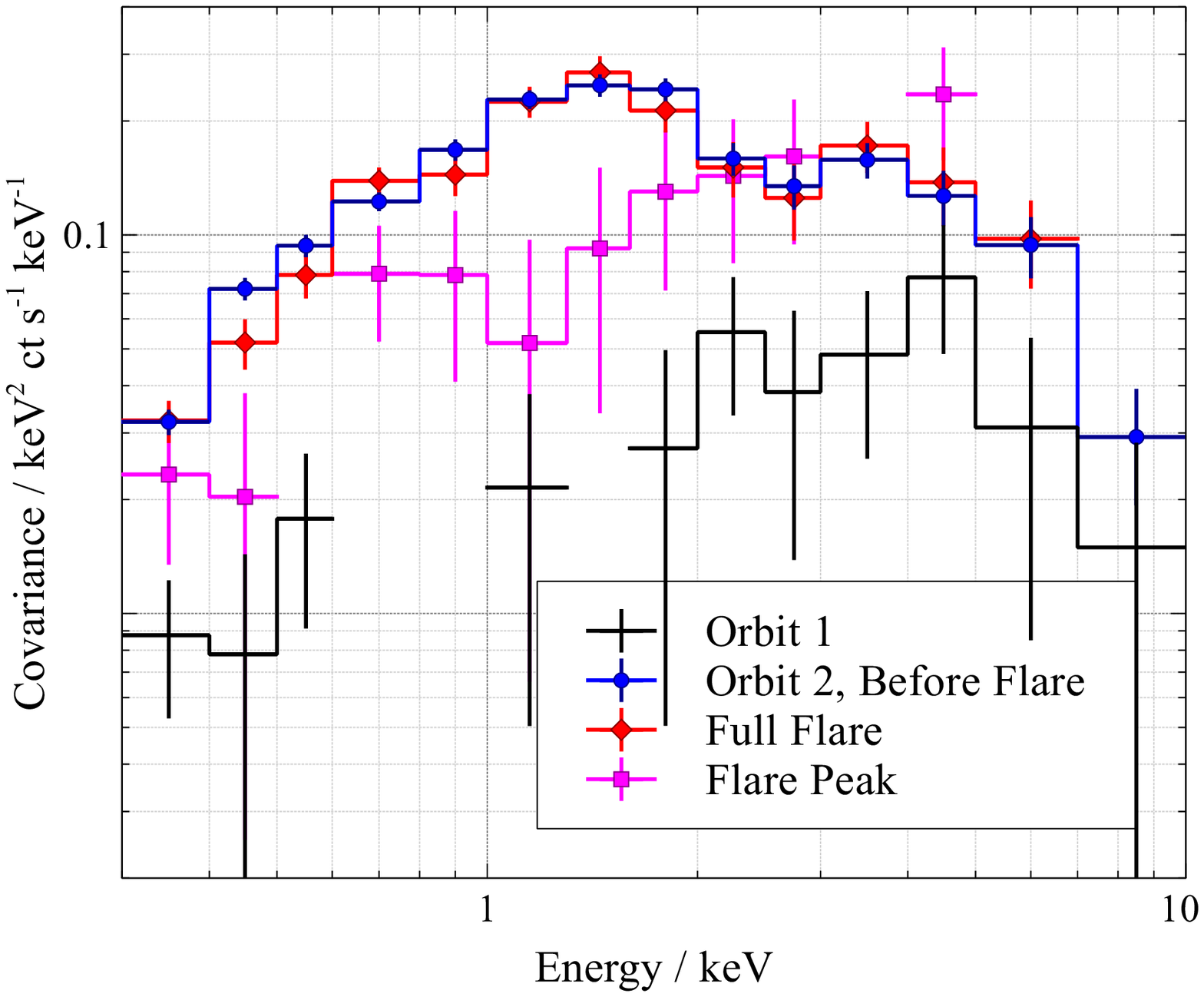}
\label{covariance.fig:high}
}
\caption[]{Covariance spectra, calculated using the EPIC MOS light curves in each energy band, showing the variable component of the spectrum over the frequency ranges in which the low and high reverberation signals were detected. The spectra are plotted in units consistent with $EF_E$ such that they represent the shape of variable spectral components.}
\label{covariance.fig}
\end{minipage}
\end{figure*}

The covariance spectra averaged over the low and high frequency ranges in which the reverberation signals were detected, $4\times 10^{-4}$ to $1.5\times 10^{-3}$\Hz\  and  $2\times 10^{-3}$ to $4\times 10^{-3}$\Hz\ respectively (the frequency range of the high frequency band is extended to increase the signal to noise, which was not possible for the lag spectra due to the loss of coherence increasing the error on the phase measurement) are shown in Fig.~\ref{covariance.fig}.

The covariance spectra are typical of those corresponding to variability predominantly arising in the power law continuum component of the X-ray emission; the variable component of the spectrum peaks between 1 and 2\keV\ where the continuum emission is most dominant and falls off towards lower energies where the continuum emission we see is reduced by Galactic absorption and to higher energies as the power law continuum falls off \citep{wilkinson_uttley, kara_1h0707}.

Over the frequency range in which the lower frequency reverberation component was found, the variation in the covariance spectrum between the time intervals before and during the flare shows the same evolution seen in the periodogram. The variability over this low frequency range rises during the second orbit before the onset of the flare, then stays high as the count rate begins to rise then dropping to slightly above the pre-flare level seen during the first orbit during the peak at the end of the flare. The shape of the variable spectral component contributing to this range of frequencies remains consistent between all of the time intervals, with some evidence of the spectrum softening (leading to a more narrowly-peaked covariance spectrum) during the peak of the flare, although due to the short length of this time segment and reduced variability, the covariance was not well measured across the full energy range.

Over the high frequency range, we initially see that during the first orbit, the variability is dominated by a harder spectral component. Simple calculations of the covariance spectrum produced by the varying normalisation of a power law continuum show that the peak is skewed to higher energies when the underlying photon index of the continuum is reduced (producing a harder spectrum). Before the flare begins during the second orbit, the variability once again increases, but now the dominant component has softened, skewing the peak of the covariance spectrum to a lower energy, now consistent with the low frequency covariance spectrum. The covariance spectrum retains this form and enhanced variability during the first part of the flare as the X-ray count rate initially increases.

In the high frequency components, however, the variability remains high right through the peak of the flare (unlike the low frequency components which had dropped to their pre-flare level during the peak). The spectral component varying at high frequencies, however, has hardened during the peak of the flare and resembles the high frequency component during the first orbit, before the flare began.

Putting together the covariance spectra and the periodograms with the shape of the lag-energy spectra in the low and high reverberation frequency ranges enables us to infer the nature of the structures within the corona and how each is evolving during the X-ray flare. Associating, via the lag-energy profiles, the low frequency variability with continuum emission arising from a corona that extends over the inner regions of the accretion disc and the high frequency variability with a collimated core of the corona, the covariance spectra from the first orbit show that the more rapidly variable continuum emission has a harder spectrum than the more slowly varying. This might be expected if the more rapidly variable emission arises from the core of the corona; being closer to the black hole, more gravitational energy is liberated from the accretion flow, accelerating the coronal particles to higher energies.

\begin{figure}
\centering
\includegraphics[width=80mm]{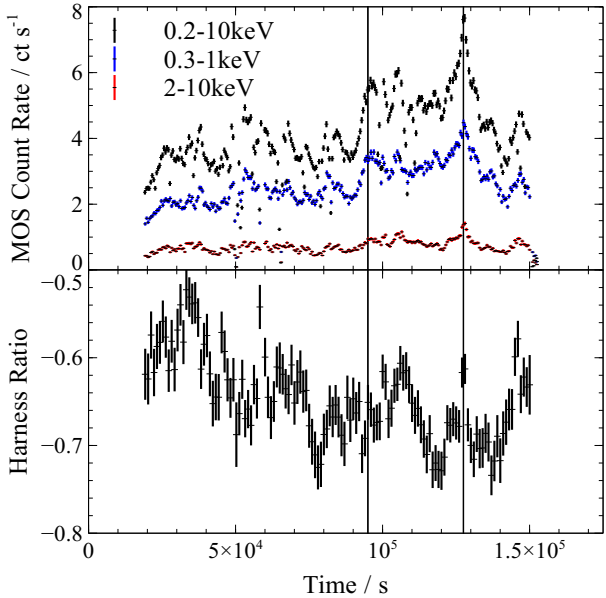}
\caption[]{The light curve of I\,Zw\,1 recorded by the EPIC MOS cameras in the full MOS band pass and in, the soft 0.3-1\keV\ band and the 2-10\keV\ band during the second orbit in which the X-ray flare was seen (\textit{top panel}). The light curves from MOS1 and MOS2 are summed. The light curve is compared to the hardness ratio calculated from the hard ($H$) and soft ($S$) bands as $(H-S)/(H+S)$ shown in the \textit{bottom panel}. The vertical lines show the start time of the full flare (the initial increase in count rate) and the sharp peak at the end of the flare.}
\label{hr.fig}
\end{figure}

It appears that the manifestation of the X-ray flare begins with an increase in variability in the extended portion of the corona but without the X-ray count rate increasing (second orbit, before the flare). The count rate from the extended corona increases some time later as the flare is seen in the X-ray light curve, though the photon index of the emitted spectrum remains constant. The variability in the extended corona has died away  prior to peak at the end of the flare (while the shape of the covariance spectrum is consistent at low frequencies, its normalisation has dropped during the peak of the flare, to the pre-peak level). At this late stage, a sharp peak is observed in the X-ray count rate as the flare passes into the collimated core of the corona; enhanced variability is seen at high frequencies coming from the harder spectral component from the collimated core of the corona. This can also be seen in the hardness ratio (Fig.~\ref{hr.fig}, compared to the light curve during the second orbit). Although the spectral evolution is clearly complex with the spectrum softening during the second orbit as the flare begins, a significant jump in the hardness ratio for a period of 2000\s\ is seen coincident with the peak of the flare; the spectrum hardens briefly as the peak of the flare emerges from the core of the corona. Although short-lived, this jump in hardness ratio is robust to the time binning that is selected (\textit{i.e.} is not an artifact of random scatter) and is seen in the hard band light curve (also shown), while not appearing in the background light curves. Full analysis of the spectral evolution of I\,Zw\,1, however, is beyond the scope of this work and will be addressed in a follow-up paper.

It is not clear how the core of the corona is affected during the initial stages of the flare. During both the pre-flare portion of the second orbit where the variability first increases and the initial period of increased X-ray count rate, the high frequency covariance spectrum has increased in magnitude and softened to have a shape consistent with the low frequency covariance spectrum. The striking similarity of the low and high frequency spectra during the pre-flare interval of the second orbit and when averaging over the full flare suggests that at the onset of the flare, the high frequencies have become dominated by the extra activity in the extended corona, washing out the activity from the core of the corona in our observations. This also explains the reverberation signature from the corona in the high frequency lag-energy spectrum being most clearly detected during the first orbit. There are insufficient data to acquire a reliable lag-energy spectrum over just the peak at the end of the flare to see if this signature once again becomes prominent during the peak and after the flare is over. While the light curves in the higher energy bands (where the count rate is lower) are affected by Poisson noise to a greater extent (particularly over the higher frequency band), the Poisson noise between energy bands is uncorrelated. This means that the shape of the covariance spectra is not expected to be impacted by noise which will simply be reflected in the error bar at each energy as it adds to the uncertainty in the measurement.


\section{Discussion}
\subsection{Structure within the corona}
\begin{figure}
\centering
\includegraphics[width=80mm]{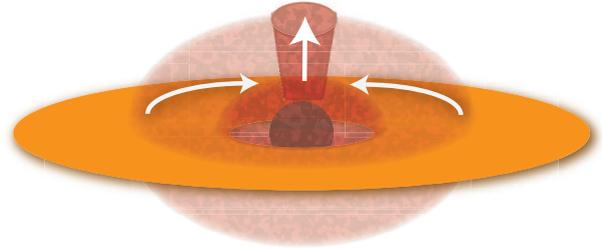}
\caption[]{The proposed structure of the corona in I\,Zw\,1, based upon the contrast between the lag-energy spectra seen at low and high frequencies. The lower frequency variability is dominated by a corona extending at low height over the surface of the accretion disc through which luminosity fluctuations propagate inwards with the accretion flow. Embedded within this is a collimated core, akin to the base of the jet. This core dominates the high frequency variability, through with fluctuations propagate upwards.}
\label{corona_core.fig}
\end{figure}

The detection of distinct reverberation lag times in the lag-frequency spectrum, 160\s\ over the frequency range $3\times 10^{-4}$\Hz\ to $1.2\times 10^{-3}$\Hz\ and 60\s\ above $1.2\times 10^{-3}$\Hz\, alongside the distinct forms of the lag-energy spectrum over each of these frequency ranges suggest that in I\,Zw\,1 the X-ray continuum is seen to arise from two distinct regions within the corona. The nature of each of these regions may be inferred from the shape of the lag-energy spectrum over each of these frequency ranges. Firstly, the lower frequency variability and its associated reverberation from the disc is seen from a corona extending over the surface of the accretion disc and, secondly, higher frequency variability is seen originating from a collimated `core' to the corona extending up the rotation axis of the black hole and perhaps resembling the base of a jet-like structure (Fig.~\ref{corona_core.fig}.

\citet{propagating_lag_paper} demonstrate that the geometry of the X-ray emitting region as well as the means by which fluctuations in luminosity fluctuate through its structure are revealed by the location of the dip signifying the earliest-responding photon energies in the lag-energy spectrum. A dip over the energy range most strongly dominated by directly observed continuum photons from the observer, from 1 to 2\keV, is na\"ively expected from simple models of X-ray reverberation from a point source located upon the spin axis of the black hole but also in the case of a corona extending, at a low height, over the surface of the accretion disc through which luminosity fluctuations propagate inwards from the outer to inner regions of the accretion disc. On the other hand, it is much more difficult to produce a dip at 3\keV\ in the lag-energy spectrum even though this has been seen almost ubiquitously in the lag-energy spectra of Seyfert galaxies \citep{kara+13}. In order to produce such a dip from propagating fluctuations within a corona whose emission is seen reverberating off an accretion disc that produces characteristic X-ray `reflection' spectra commonly employed in the spectral modelling of AGN, \citet{propagating_lag_paper} show that it is necessary for fluctuations to propagate up a vertically collimated corona at a relatively low velocity ($\sim 0.01c$). The presence of a rapidly varying collimated core to the corona suggests that the lag-energy profile obtained over the lower frequency reverberation dip should be interpreted as an extended corona rather than as a centrally located point source, since it seems unlikely that there is a slowly varying point within the rapidly varying vertically extended core. While this may not turn out to be a unique interpretation of the data, thus far the models of \citet{propagating_lag_paper} provide the most plausible explanation for the dip at 3\keV\ in lag-energy spectra of Seyfert galaxies and how the low frequency lag-frequency spectrum of I\,Zw\,1 can deviate from this with a dip between 1 and 2\keV, while the 3\keV\ dip is simultaneously seen at higher frequencies.

The mass of the black hole in I\,Zw\,1 has been found to be $2.8_{-0.7}^{+0.6} \times 10^7$\Msun, inferred from the width of the broad H$\beta$ line \citep[][where I\,Zw\,1 is tabulated by its alternative identifier, PG\,0050+124]{vestergaard+06}, corresponding to a characteristic light-crossing time over a gravitational radius of $GM/c^3=140$\s. Over the lower frequency range, the reverberation timescale was measured to be 160\s. \citet{lag_spectra_paper} show that the measured reverberation lag can be diluted by up to 50 or even 75 per cent in extreme cases owing to the fact that we do not measure energy bands comprised solely of continuum and reflected photons, rather the soft, reflection band has some contribution from reflected emission and \textit{vice versa}. The measured time lag is consistent with that measured in other Seyfert galaxies \citep{demarco+2012,kara+2016} insofar as the extended portion of the corona extends just $2\sim 3$\rg\ above the plane of the accretion disc. Such a trivial analysis cannot, however, be applied to estimate the scale height of the extended core of the corona since the maximal time lag is measured not between the reflection-dominated soft excess and the continuum, rather between the soft excess and the 3\keV\ dip which is dominated by the extremely redshifted iron K$\alpha$ emission from the inner regions of the disc.

The relative frequencies at which the reverberation signatures of these two coronal structures are seen are as expected. The lower frequency variability is associated with the extended corona that spans the inner regions of the accretion disc. If the emission from this corona is governed by underlying variations in the accretion flow, for instance turbulence in the disc, the accretion of over- and under-dense parts of the disc or the accumulation of magnetic flux, one might expect that variability frequencies are limited by either the relatively long orbital timescale at the radii over which coronal X-ray emission is significant or the viscous timescale upon which material flows inwards. The characteristic timescale of an orbit in the equatorial plane of the Kerr spacetime (maximally rotating with spin parameter $a=0.998\,GM/c^2$) at radius 5\rg\ is $77\,GM/c^3$ which, taking the lower limit black hole mass to be $2\times 10^7$\Msun, corresponds to a timescale of 8\ks\ or a characteristic frequency of $10^4$\Hz. At the innermost stable circular orbit ($r_\mathrm{ISCO}=1.235$\rg), the timescale is $15\,GM/c^3$ or 1.5\ks and the characteristic frequency is $7\times 10^4$\Hz. It is thus plausible that the variability in the extended corona is generated across the inner regions of the underlying accretion disc.

On the other hand, a collimated corona is likely powered from the inner regions of the accretion flow where the orbital period is much shorter or from magnetic field configurations within the inner regions, hence at higher frequencies, only the collimated core and not the extended portion of the corona contributes to the variability that is seen. The variability that is attributed to the collimated core of the corona is seen on timescales shorter than the orbital timescale around the innermost stable orbit and it seems that more rapid variability is being induced by either small scale turbulence in the accretion flow or magnetic field structure that is short-lived and not smoothed out by the orbital motion or is due to turbulence in the black hole magnetosphere close to the event horizon and the spin axis.

One might associate the extended portion of the corona with magnetic field lines that thread the accretion disc and drive the accretion flow by providing the necessary stresses to transfer angular momentum away from material on inner orbits, allowing it to flow inwards, towards the black hole. These fields may provide these stresses through either the magneto-rotational instability  \citep{balbus+91,balbus+98} or through magnetic braking by large-scale field lines \citep{blandford_payne}. In regions where a sufficient flux density is able to build up above the accretion disc, perhaps as it accretes inwards, as has been demonstrated in GRMHD simulations of both thick and strongly-magnetised thin discs \citep{mckinney+2012,avara+2016}, it may be able to accelerate the particles (perhaps when field lines undergo reconnection events) that are able to Comptonise thermal photons from the disc and in so doing produce the X-ray continuum.

The collimated core to the corona is likely driven by field lines either in the black hole magnetosphere or anchored to the inner accretion disc with the structure collimated as field lines are wrapped around the rotation axis by the orbital motion and by frame-dragging due to the rotating black hole. Such a configuration may allow energy to be extracted from the spin of the black hole in order to power the corona \citep{blandford_znajek}. I\,Zw\,1 is known to be a comparatively radio-quiet source, spatially unresolved with physical size $\le 0.38$\kpc\ and with a steep radio spectrum indicative of optically thin synchrotron emission \citep{kukula+98}. The low luminosity and compact nature of the radio source suggest that while there may be particle motion (or at least the propagation of luminosity fluctuations) up a collimated core, these particles are not able to escape to form extended radio jets, whether because the velocity is insufficient to escape the gravitational field of the black hole as in the aborted jet model of \citet{ghisellini+04} or because the magnetic field configuration is such that the would-be jet power is dissipated in an X-ray emitting corona.

\citet{propagating_lag_paper} suggest that both of these coronal components must be present in order to self-consistently explain the presence of both `hard lags' seen at low frequencies due to a propagating fluctuations model \citep{kotov+2001,arevalo+2006} and the detection of reverberation time lags. In many cases, the hard lag is seen at low frequencies, while the more rapid variability from the collimated inner part of the corona drives the reverberation signal that is seen at higher frequencies. The lag-energy spectrum corresponding to reverberation from the more slowly varying extended portion of the corona is not seen directly since over high frequency ranges, the more rapidly varying core of the corona is dominant while at lower frequencies, the dominant effect is the hard lag due to the gradient in continuum photon index produced as the fluctuation propagates inwards through the disc/corona.

In I\,Zw\,1, it appears that in addition to the hard lag at low frequencies and the reverberation from the collimated core of the corona at high frequencies, there is an intermediate frequency range over which it is possible to observe reverberation from the extended outer part of the corona. This is perhaps aided by the larger mass of the black hole in I\,Zw\,1; $2.8\times 10^7$\Msun  compared to $\sim 10^6$\Msun\ in NLS1 galaxies such as 1H\,0707$-$495 in which reverberation has been studied in most detail. This allows for a greater absolute difference in variability timescale between the extended and inner parts of the corona and by a strong source of variability driving reverberation from the extended corona, meaning that this reverberation becomes the dominant process over the intermediate frequency range. If the stochastic variability over specific radii within the corona is stronger than that which propagates inwards, or if the stochastic variability generated at specific radii is more rapid than the inward propagation, reverberation will dominate over the hard, propagation lag over the frequency range in question.

\citet{gallo_1zw1_2} previously found evidence for multiple emission components within the corona of I\,Zw\,1 during 2002 and 2005 observations. It was found that in order to fit the spectrum over the full \textit{XMM-Newton} bandpass, it is necessary to include two continuum components. Fitting a hard power law continuum over the 2-10\keV\ energy band left residuals representing a softer continuum of which the normalisation was highly erratic, although its shape remained constant. In addition to the continuum, both neutral and ionised absorption is seen in I\,Zw\,1, constant over the course of the observations \citep{costantini+07} as well as an iron K$\alpha$ fluorescence line originating a few tens of gravitational radii from the black hole, determined by the maximal redshift seen in the line \citep{gallo_1zw1_1}.

\citet{gallo_1zw1_2} found two distinct modes in the variability of the hard continuum component in the 2005 observation. The transition between these was marked by a sharp dip in X-ray flux. Before the dip, both the slope and normalisation of this emission component were seen to vary, while after variability was only detected in its normalisation. Similar transitions between modes of variability were seen during the 2002 observation where a hard X-ray flare was found to originate from closer to the black hole from both time lags between hard and soft X-rays and from more extreme redshifts in the iron K$\alpha$ line being seen during the flare \citep{gallo_1zw1_1}. The softer continuum component was inferred to originate from an optically thin plasma over the disc owing to the modest redshifts in the iron line, indicating reflection around 10\rg\ from the black hole while a central component was found to be responsible for the harder continuum as well as rapid events including flares and dips. In this work, we are able to directly detect the presence of these two coronal components and determine the location and geometry of each from X-ray spectral timing and demonstrate that these two structures co-exist within the corona.

The detection of two principal components in PCA of the spectral variability, each corresponding to the change in slope of a power law-like continuum supports the existence of two coronal components with variability uncorrelated between the two. Note that variability in the collimated core may still be modulated by that in the extended corona as fluctuations propagate inwards through the accretion flow. The detection of the second, uncorrelated principal component merely shows that there as a strong contribution of additional, uncorrelated variability from the collimated core, driven by turbulence and variations arising from the very inner parts of the accretion flow.

Comparing the covariance spectra over the low and high frequency bands from the first orbit of the observations suggests that the more rapidly variable continuum emission from the core of the corona possesses a harder spectrum than the continuum emission from the extended portion of the corona above the accretion disc. Such a distinction might be expected if one assumes that more energy is liberated from the accretion flow in the inner regions of the disc and corona where the material has fallen deeper into the gravitational potential or where magnetic flux has accreted and become more concentrated. If the coronal continuum emission is produced by the Comptonisation of black body seed photons from the disc, a harder spectrum indicates either a higher characteristic electron temperature in the inner regions of the corona as greater magnetic flux density accelerates the particles to greater energies or it indicates a greater optical depth through the corona to the seed photons scattering from the inner regions. Without measurements of the high frequency X-ray variability above 10\keV, however, it is not possible to make this distinction. A harder continuum originating from the inner regions of the corona supports propagating fluctuation models of the low frequency X-ray variability \citep{kotov+2001,arevalo+2006} where the propagation of mass accretion rate fluctuations inwards through the accretion disc modulates the softer X-ray emission before the harder emission from the inner regions.

\subsection{Coronal flaring mechanisms}

Over the course of the 2015 observations, evidence is seen for the separate evolution of the detected structures within the corona. This evolution is associated with a flare in the X-ray flux during the second orbit of the observation. The count rate initially increased by 15 per cent from the mean level up until this point, then stabilised at this increased level for 29\ks\ before increasing a further 44 per cent during a sharp peak lasting 17\ks\ at the end of the flare. The X-ray count rate then returned to its pre-flare level.

During the first orbit, before the onset of the flare, only the higher frequency reverberation signature is seen which, through its lag-energy profile, is associated with upward propagation through the collimated core of the corona. The power spectral density (PSD) of the variability falls off relatively steeply as approximately $f^{-2}$ and, during the first orbit, there is insufficient variability in the emission from the portion of the corona extended over the accretion disc for a significant reverberation signal to be detected from it. The narrower profile and the peak at lower X-ray energies in the low frequency covariance spectrum compared to the high frequency covariance spectrum shows, however, that the emission dominating the weaker low frequency variability has a softer spectrum than that dominating the high frequency variability, thus that it arises from a different part of the corona.

During the second orbit of the observation, however, evolution in the extended portion of the corona, identified from its distinct lag-energy profile detected over a lower frequency range, becomes apparent even before the increase in X-ray count rate is seen. Right from the beginning of the second orbit, the variability was found to increase substantially. The PSD was found to have flattened, now falling off as approximately $f^{-1.1}$. With significantly more variability over the lower frequency range associated with the variability in the emission from the extended corona, the signature of reverberation from this corona became clearly visible in the lag-energy spectrum. It appears that the first stage in the coronal flaring mechanism is an increase in activity within the corona above the disc, increasing the variability with the magnetic field structure above the disc perhaps becoming more turbulent. While the extended corona becomes more turbulent, it does not initially produce an increase in the X-ray count rate. If the coronal emission is due to Comptonisation seed photons from the disc, this means that the combination of the seed photon flux from the disc and the cross section of the corona for scattering the seed photons remains unchanged during these early stages.

The increase in X-ray count rate follows some time after the initial increase in turbulent activity within the extended corona. The X-ray count rate first increases $70\sim 110$\ks\ after the variability is seen to increase (with the uncertainty arising due to the time interval between the first and second \textit{XMM-Newton} orbits and the PSD having increased some time between the orbits of the observation). The variability remains high after the initial increase in count rate (after which the count rate stabilised for 29\ks) with the PSD falling off as only $f^{-0.8}$. The shape and location of the peak in the covariance spectra reveals that both the low and high frequency variability are dominated by the softer continuum emission from the extended part of the corona. This could well be the result of an accumulation or generation of additional magnetic flux above the disc. The enhanced flux density results into more frequent but still stochastic reconnection events \citep[\textit{e.g.}][]{merloni_fabian} increasing both the count rate and the variability.

The PSDs of AGN typically exhibit a broken power law profile, transitioning from $f^{-1}$ to $f^{-2}$ at some break frequency \citep{mchardy+2004, mchardy+2007}. This break frequency has been found to scale with both the mass of the black hole, breaking at lower frequencies for greater mass, and with bolometric luminosity with brighter sources possessing a higher break frequency \citep{mchardy+2006}. That we find the best-fitting power law to the periodogram vary from $f^{-2}$ to $f^{-1.1}$ and $f^{-0.8}$ during the run-up to and initial onset of the flare is highly suggestive that we were witnessing a shift in the break frequency in I\,Zw\,1. Assuming a black hole mass of $2.8\times 10^7$\Msun\ and bolometric luminosity $5\times 10^{45}$\ergps\ \citep{martinez-paredes+2017}, the relation of \citet{mchardy+2006} predicts a break frequency $1\times 10^{-4}$\Hz; consistent with the periodogram falling off as $f^{-2}$ above this frequency before the flare (and during the peak). During the peak, however, it is possible that the break frequency has shifted above $10^{-2}$\Hz\ and we do not see the $f^{-2}$ high frequency PSD, suggesting that the break frequency or cut-off in the PSD in AGN is associated with the structure of the corona (of course, still scaled accordingly for the mass of the black hole) and that during some transient events such as this flare, it can change on short timescales. It is also possible that the break frequency remains constant, around $10^{-4}$\Hz, and it is the sharpness of the bend in the PSD that changes, with the power spectrum becoming flatter during the run-up to and initial phases of the flare when the softer spectral component arising from the extended portion of the corona becomes more variable and falling off more steeply when the harder spectral component from the core of the component is driving the variability. Such behaviour was found by \citet{grinberg+2014} in the variation of the power spectrum during the high soft state of the high mass X-ray binary Cygnus X-1 as the spectral shape changes. The power spectrum was found to bend more sharply when the X-ray spectrum was harder.

A broken power law is likely an oversimplification of the shape of the PSD, for instance \citet{mchardy+2007} find that the power spectrum of the narrow line Seyfert 1 galaxy Ark~564 can be modeled as the sum of two Lorentzian components with a distinct time lag between the variability described by each of these, suggesting that the variability in the X-ray emission in this source is generated in two localised regions. It is plausible that the variation in the observed in the PSD of I\,Zw\,1 over the course of the flare could be described in similar terms, with a constant low frequency component and the additional variability seen at higher frequencies in the PSD described by a second, variable narrow band component.

To test this hypothesis, the periodogram in each of the time intervals was fit as the sum of a low and a high frequency Lorentzian component. Initially, the amplitude, centroid and width of the low frequency Lorentzian was tied between the time intervals and the high frequency component allowed to vary in amplitude (but the centroid and width were tied). This did not provide a good description of the data; the additional variability power seen at low frequencies during the pre-flare period of the second orbit and the start of the flare was not accounted for ($\chi^2/\nu = 7.1$). Allowing the amplitude of the low frequency Lorentzian component to also vary between the time intervals improves the fit to the measured periodograms but is still not formally acceptable ($\chi^2/\nu = 2.7$). The additional variability at the lowest frequencies is now accounted for but there is excess variability not well modeled between $2\times 10^{-4}$ and $10^{-3}$\Hz\ and in the tail at the highest frequencies sampled. We note however that the quality of the fit is similar to that of a simple power law PSD in each time interval ($\chi^2/\nu = 2.0$) where similar residuals are seen. While it seems plausible from the measured periodograms that the evolution of the PSD of I\,Zw\,1 during the flare can be attributed to the changing amplitudes of two narrow-band components, it is clear that the variability is more complex than simply the sum of two Lorentzian components changing in amplitude  and the periodogram that can be measured by \textit{XMM-Newton} over the limited time intervals in question is not sufficient in frequency range or signal-to-noise to shed much more light on the components at play and further constrain these models.

As the X-ray emission rises into the peak at the end of the flare, the low frequency variability associated with the emission from the extended portion of the corona has died away. The high frequency variability, however, remains high and the high frequency covariance spectrum once again shows the profile associated with the harder X-ray emission from the core of the corona. We see clear evidence for the propagation of the X-ray flare through the corona, beginning in the accretion disc and increasing the X-ray count rate from the parts of the corona extended over the disc. This initial increase in count rate is relatively modest and sustained over a prolonged period as the count rate flattens off at the increased level for 29\ks. The flaring passes inwards through the extended corona to the core. The low frequency variability dies away but a much greater and more rapid increase in X-ray flux is witnessed from the inner regions associated with harder X-ray emission that is variable at higher frequencies. Such rapid X-ray peaks have previously been associated with the collimated core of the corona in the narrow line Seyfert 1 galaxy Markarian 335, during which portions of the corona were launched upwards, reminiscent of the base of a jet that has failed to launch \citep{mrk335_corona_paper, mrk335_flare_paper}.

A shift in the reverberation signal to lower frequencies along with an increase in the reverberation lag time was seen as the X-ray flux increased from the NLS1 galaxy IRAS\,13224$-$3809 \citep{kara_iras_lags}. An increase in time lag along with a shift to lower frequencies can be interpreted as an increase in the scale-height of the corona above the disc and further from the black hole \citep{lag_spectra_paper,cackett_ngc4151}. In this case, however, two distinct reverberation signatures are seen with different lag-energy profiles suggesting the coexistence of the extended and collimated structures in the corona rather than the motion or expansion of a single corona further from the black hole.

While the rms-flux relation has been well established in accreting black holes, with greater variability seen as the X-ray flux increases \citep{uttley+2001,vaughan+2003}, we now see that there is hysteresis associated with the structure and variability of the corona on short timescales during transient events such as X-ray flares. Within this single source, there is not a one-to-one mapping between the X-ray count rate and the rms variability and structure of the corona; rather there is evolution wherein the variability increase precedes the flare in the X-ray count rate and the flare passes from the extended portion of the corona to the collimated core.


\section{Conclusions}
X-ray timing analysis of the narrow line Seyfert 1 galaxy I~Zwicky~1 suggests X-ray reverberation originating from distinct regions within the corona.

The reverberation time lag between correlated variability in the 0.3-1.0\keV\ and 1.0-4.0\keV\ light curves was found to shorten from 160\s\ over the frequency range $3\times 10^{-4}$\Hz\ to $1.2\times 10^{-3}$\Hz\ to 60\s\ above $1.2\times 10^{-3}$\Hz. The lag-energy spectrum over each reveals the geometry of the component of the corona that dominates the variability over each of these frequency ranges when compared to theoretical models of X-ray reverberation derived from general relativistic ray tracing simulations.

The lag-energy spectrum over the lower frequency range shows a dip corresponding to the earliest arriving photons at $(1.62^{+0.03}_{-0.14})$\keV, with later arrival of photons below this in the reflected soft excess and in the iron K$\alpha$ line around 6\keV, indicating the reverberation of photons originating from a corona extended at a low height over the surface of the disc. On the other hand, the lag-energy spectrum over the higher of these frequency ranges shows a suggestive dip at $(3.7^{+1}_{-0.4})$\keV, typical of Seyfert galaxies. The 3\keV\ dip can be produced in addition to the 1-2\keV\ dip at lower variability frequencies if this more rapid variability is dominated by X-rays originating from a collimated core to the corona through which luminosity fluctuations propagate upwards, reminiscent of the base of a jet (although I\,Zw\,1 is a radio quiet source).

Principal component analysis of the X-ray spectral variability finds three significant eigenvectors corresponding to the change in normalisation of the spectrum alongside the uncorrelated variation in the spectral slope of two power law continua, supporting the interpretation of distinct structures within the corona dominating each of the low and high frequency variability.

Analysis of the X-ray variability and reverberation lag over the course of the observations reveals the evolution of the corona during an X-ray flare. The low frequency variability in the emission from the extended portion of the corona was found to increase even before the X-ray count rate increased. The variability in the emission from the extended corona remained high as the count rate first increased. A second, more sudden increase in the X-ray count rate was seen as the flare passed into the collimated core of the corona. At this time, the low frequency variability from the extended corona had died away, but the high frequency variability in the harder spectral component remained high.

These observations show the power of X-ray timing analysis from long, continuous X-ray observations of AGN to not just constrain the location and geometry of the corona, but discover structures within the corona as well as how these evolve as the X-ray count rate varies. By understanding the details of the corona, its structure and its evolution, we will be able to learn how through magnetic and other processes energy is liberated from material accreting onto supermassive black holes to power some of the most luminous objects we see in the Universe.

\section*{Acknowledgements}
DRW is supported by NASA through Einstein Postdoctoral Fellowship grant number PF6-170160, awarded by the \textit{Chandra} X-ray Center, operated by the Smithsonian Astrophysical Observatory for NASA under contract NAS8-03060. CVS acknowledges support from NOVA (Nederlandse Onderzoekschool voor Astronomie). The Space Research Organization of the Netherlands is supported financially by NWO, the Netherlands Organization for Scientific Research. EC is partially supported by the NWO-Vidi grant number 639.042.525. WNB acknowledges support through Space Telescope Science Institute grant HST-GO-13811.004-A. Thanks must go to Margherita Giustini for useful discussions during the analysis of these data. We thank the anonymous referee for their useful feedback on the initial version of this manuscript.

\vspace{-0.1cm}
\bibliographystyle{mnras}
\bibliography{agn}

\begin{thebibliography}{}
\makeatletter
\relax
\def\mn@urlcharsother{\let\do\@makeother \do\$\do\&\do\#\do\^\do\_\do\%\do\~}
\def\mn@doi{\begingroup\mn@urlcharsother \@ifnextchar [ {\mn@doi@}
  {\mn@doi@[]}}
\def\mn@doi@[#1]#2{\def\@tempa{#1}\ifx\@tempa\@empty \href
  {http://dx.doi.org/#2} {doi:#2}\else \href {http://dx.doi.org/#2} {#1}\fi
  \endgroup}
\def\mn@eprint#1#2{\mn@eprint@#1:#2::\@nil}
\def\mn@eprint@arXiv#1{\href {http://arxiv.org/abs/#1} {{\tt arXiv:#1}}}
\def\mn@eprint@dblp#1{\href {http://dblp.uni-trier.de/rec/bibtex/#1.xml}
  {dblp:#1}}
\def\mn@eprint@#1:#2:#3:#4\@nil{\def\@tempa {#1}\def\@tempb {#2}\def\@tempc
  {#3}\ifx \@tempc \@empty \let \@tempc \@tempb \let \@tempb \@tempa \fi \ifx
  \@tempb \@empty \def\@tempb {arXiv}\fi \@ifundefined
  {mn@eprint@\@tempb}{\@tempb:\@tempc}{\expandafter \expandafter \csname
  mn@eprint@\@tempb\endcsname \expandafter{\@tempc}}}

\bibitem[\protect\citeauthoryear{{Ar{\'e}valo} \& {Uttley}}{{Ar{\'e}valo} \&
  {Uttley}}{2006}]{arevalo+2006}
{Ar{\'e}valo} P.,  {Uttley} P.,  2006, \mn@doi [\mnras]
  {10.1111/j.1365-2966.2006.09989.x}, \href
  {http://adsabs.harvard.edu/abs/2006MNRAS.367..801A} {367, 801}

\bibitem[\protect\citeauthoryear{{Avara}, {McKinney}  \& {Reynolds}}{{Avara}
  et~al.}{2016}]{avara+2016}
{Avara} M.~J.,  {McKinney} J.~C.,   {Reynolds} C.~S.,  2016, \mn@doi [\mnras]
  {10.1093/mnras/stw1643}, \href
  {http://adsabs.harvard.edu/abs/2016MNRAS.462..636A} {462, 636}

\bibitem[\protect\citeauthoryear{{Balbus} \& {Hawley}}{{Balbus} \&
  {Hawley}}{1991}]{balbus+91}
{Balbus} S.~A.,  {Hawley} J.~F.,  1991, \mn@doi [\apj] {10.1086/170270}, \href
  {http://adsabs.harvard.edu/abs/1991ApJ...376..214B} {376, 214}

\bibitem[\protect\citeauthoryear{{Balbus} \& {Hawley}}{{Balbus} \&
  {Hawley}}{1998}]{balbus+98}
{Balbus} S.~A.,  {Hawley} J.~F.,  1998, \mn@doi [Reviews of Modern Physics]
  {10.1103/RevModPhys.70.1}, \href
  {http://adsabs.harvard.edu/abs/1998RvMP...70....1B} {70, 1}

\bibitem[\protect\citeauthoryear{{Blandford} \& {Payne}}{{Blandford} \&
  {Payne}}{1982}]{blandford_payne}
{Blandford} R.~D.,  {Payne} D.~G.,  1982, \mnras, \href
  {http://adsabs.harvard.edu/abs/1982MNRAS.199..883B} {199, 883}

\bibitem[\protect\citeauthoryear{{Blandford} \& {Znajek}}{{Blandford} \&
  {Znajek}}{1977}]{blandford_znajek}
{Blandford} R.~D.,  {Znajek} R.~L.,  1977, \mnras, \href
  {http://adsabs.harvard.edu/abs/1977MNRAS.179..433B} {179, 433}

\bibitem[\protect\citeauthoryear{{Boller}, {Brandt}  \& {Fink}}{{Boller}
  et~al.}{1996}]{boller+96}
{Boller} T.,  {Brandt} W.~N.,   {Fink} H.,  1996, \aap, \href
  {http://adsabs.harvard.edu/abs/1996A\%26A...305...53B} {305, 53}

\bibitem[\protect\citeauthoryear{{Cackett}, {Zoghbi}, {Reynolds}, {Fabian},
  {Kara}, {Uttley}  \& {Wilkins}}{{Cackett} et~al.}{2014}]{cackett_ngc4151}
{Cackett} E.~M.,  {Zoghbi} A.,  {Reynolds} C.,  {Fabian} A.~C.,  {Kara} E.,
  {Uttley} P.,   {Wilkins} D.~R.,  2014, \mn@doi [\mnras]
  {10.1093/mnras/stt2424}, \href
  {http://adsabs.harvard.edu/abs/2014MNRAS.438.2980C} {438, 2980}

\bibitem[\protect\citeauthoryear{{Chainakun}, {Young}  \& {Kara}}{{Chainakun}
  et~al.}{2016}]{chainakun+2016}
{Chainakun} P.,  {Young} A.~J.,   {Kara} E.,  2016, \mn@doi [\mnras]
  {10.1093/mnras/stw1105}, \href
  {http://adsabs.harvard.edu/abs/2016MNRAS.460.3076C} {460, 3076}

\bibitem[\protect\citeauthoryear{{Costantini}, {Gallo}, {Brandt}, {Fabian}  \&
  {Boller}}{{Costantini} et~al.}{2007}]{costantini+07}
{Costantini} E.,  {Gallo} L.~C.,  {Brandt} W.~N.,  {Fabian} A.~C.,   {Boller}
  T.,  2007, \mn@doi [\mnras] {10.1111/j.1365-2966.2007.11646.x}, \href
  {http://adsabs.harvard.edu/abs/2007MNRAS.378..873C} {378, 873}

\bibitem[\protect\citeauthoryear{{De Marco}, {Ponti}, {Cappi}, {Dadina},
  {Uttley}, {Cackett}, {Fabian}  \& {Miniutti}}{{De Marco}
  et~al.}{2013}]{demarco+2012}
{De Marco} B.,  {Ponti} G.,  {Cappi} M.,  {Dadina} M.,  {Uttley} P.,  {Cackett}
  E.~M.,  {Fabian} A.~C.,   {Miniutti} G.,  2013, \mn@doi [\mnras]
  {10.1093/mnras/stt339}, \href
  {http://adsabs.harvard.edu/abs/2013MNRAS.431.2441D} {431, 2441}

\bibitem[\protect\citeauthoryear{{Epitropakis} \& {Papadakis}}{{Epitropakis} \&
  {Papadakis}}{2016}]{epit+16}
{Epitropakis} A.,  {Papadakis} I.~E.,  2016, \mn@doi [\aap]
  {10.1051/0004-6361/201527665}, \href
  {http://adsabs.harvard.edu/abs/2016A\%26A...591A.113E} {591, A113}

\bibitem[\protect\citeauthoryear{{Fabian}, {Rees}, {Stella}  \&
  {White}}{{Fabian} et~al.}{1989}]{fabian+89}
{Fabian} A.~C.,  {Rees} M.~J.,  {Stella} L.,   {White} N.~E.,  1989, \mnras,
  \href {http://ukads.nottingham.ac.uk/abs/1989MNRAS.238..729F} {238, 729}

\bibitem[\protect\citeauthoryear{{Fabian} et~al.,}{{Fabian}
  et~al.}{2009}]{fabian+09}
{Fabian} A.~C.,  et~al., 2009, \mn@doi [\nat] {10.1038/nature08007}, \href
  {http://adsabs.harvard.edu/abs/2009Natur.459..540F} {459, 540}

\bibitem[\protect\citeauthoryear{{Gallo}}{{Gallo}}{2006}]{gallo_nls1}
{Gallo} L.~C.,  2006, \mn@doi [\mnras] {10.1111/j.1365-2966.2006.10137.x},
  \href {http://adsabs.harvard.edu/abs/2006MNRAS.368..479G} {368, 479}

\bibitem[\protect\citeauthoryear{{Gallo}, {Brandt}, {Costantini}, {Fabian},
  {Iwasawa}  \& {Papadakis}}{{Gallo} et~al.}{2007a}]{gallo_1zw1_1}
{Gallo} L.~C.,  {Brandt} W.~N.,  {Costantini} E.,  {Fabian} A.~C.,  {Iwasawa}
  K.,   {Papadakis} I.~E.,  2007a, \mn@doi [\mnras]
  {10.1111/j.1365-2966.2007.11601.x}, \href
  {http://adsabs.harvard.edu/abs/2007MNRAS.377..391G} {377, 391}

\bibitem[\protect\citeauthoryear{{Gallo}, {Brandt}, {Costantini}  \&
  {Fabian}}{{Gallo} et~al.}{2007b}]{gallo_1zw1_2}
{Gallo} L.~C.,  {Brandt} W.~N.,  {Costantini} E.,   {Fabian} A.~C.,  2007b,
  \mn@doi [\mnras] {10.1111/j.1365-2966.2007.11701.x}, \href
  {http://adsabs.harvard.edu/abs/2007MNRAS.377.1375G} {377, 1375}

\bibitem[\protect\citeauthoryear{{Ghisellini}, {Haardt}  \&
  {Matt}}{{Ghisellini} et~al.}{2004}]{ghisellini+04}
{Ghisellini} G.,  {Haardt} F.,   {Matt} G.,  2004, \mn@doi [\aap]
  {10.1051/0004-6361:20031562}, \href
  {http://adsabs.harvard.edu/abs/2004A\%26A...413..535G} {413, 535}

\bibitem[\protect\citeauthoryear{{Grinberg} et~al.,}{{Grinberg}
  et~al.}{2014}]{grinberg+2014}
{Grinberg} V.,  et~al., 2014, \mn@doi [\aap] {10.1051/0004-6361/201322969},
  \href {http://adsabs.harvard.edu/abs/2014A%26A...565A...1G} {565, A1}

\bibitem[\protect\citeauthoryear{{Kara}, {Fabian}, {Cackett}, {Steiner},
  {Uttley}, {Wilkins}  \& {Zoghbi}}{{Kara} et~al.}{2013a}]{kara_1h0707}
{Kara} E.,  {Fabian} A.~C.,  {Cackett} E.~M.,  {Steiner} J.~F.,  {Uttley} P.,
  {Wilkins} D.~R.,   {Zoghbi} A.,  2013a, \mn@doi [\mnras]
  {10.1093/mnras/sts155}, \href
  {http://adsabs.harvard.edu/abs/2013MNRAS.428.2795K} {428, 2795}

\bibitem[\protect\citeauthoryear{{Kara}, {Fabian}, {Cackett}, {Miniutti}  \&
  {Uttley}}{{Kara} et~al.}{2013b}]{kara_iras_lags}
{Kara} E.,  {Fabian} A.~C.,  {Cackett} E.~M.,  {Miniutti} G.,   {Uttley} P.,
  2013b, \mn@doi [\mnras] {10.1093/mnras/stt024}, \href
  {http://adsabs.harvard.edu/abs/2013MNRAS.430.1408K} {430, 1408}

\bibitem[\protect\citeauthoryear{{Kara}, {Fabian}, {Cackett}, {Uttley},
  {Wilkins}  \& {Zoghbi}}{{Kara} et~al.}{2013c}]{kara+13}
{Kara} E.,  {Fabian} A.~C.,  {Cackett} E.~M.,  {Uttley} P.,  {Wilkins} D.~R.,
  {Zoghbi} A.,  2013c, \mn@doi [\mnras] {10.1093/mnras/stt1055}, \href
  {http://adsabs.harvard.edu/abs/2013MNRAS.434.1129K} {434, 1129}

\bibitem[\protect\citeauthoryear{{Kara}, {Alston}, {Fabian}, {Cackett},
  {Uttley}, {Reynolds}  \& {Zoghbi}}{{Kara} et~al.}{2016}]{kara+2016}
{Kara} E.,  {Alston} W.~N.,  {Fabian} A.~C.,  {Cackett} E.~M.,  {Uttley} P.,
  {Reynolds} C.~S.,   {Zoghbi} A.,  2016, \mn@doi [\mnras]
  {10.1093/mnras/stw1695}, \href
  {http://adsabs.harvard.edu/abs/2016MNRAS.462..511K} {462, 511}

\bibitem[\protect\citeauthoryear{{Kotov}, {Churazov}  \& {Gilfanov}}{{Kotov}
  et~al.}{2001}]{kotov+2001}
{Kotov} O.,  {Churazov} E.,   {Gilfanov} M.,  2001, \mn@doi [\mnras]
  {10.1046/j.1365-8711.2001.04769.x}, \href
  {http://adsabs.harvard.edu/abs/2001MNRAS.327..799K} {327, 799}

\bibitem[\protect\citeauthoryear{{Kukula}, {Dunlop}, {Hughes}  \&
  {Rawlings}}{{Kukula} et~al.}{1998}]{kukula+98}
{Kukula} M.~J.,  {Dunlop} J.~S.,  {Hughes} D.~H.,   {Rawlings} S.,  1998,
  \mn@doi [\mnras] {10.1046/j.1365-8711.1998.01481.x}, \href
  {http://adsabs.harvard.edu/abs/1998MNRAS.297..366K} {297, 366}

\bibitem[\protect\citeauthoryear{{Mart\'{\i}nez-Paredes}
  et~al.,}{{Mart\'{\i}nez-Paredes} et~al.}{2017}]{martinez-paredes+2017}
{Mart\'{\i}nez-Paredes} M.,  et~al., 2017, \mn@doi [\mnras]
  {10.1093/mnras/stx307}, \href
  {http://adsabs.harvard.edu/abs/2017MNRAS.468....2M} {468, 2}

\bibitem[\protect\citeauthoryear{{McHardy}, {Papadakis}, {Uttley}, {Page}  \&
  {Mason}}{{McHardy} et~al.}{2004}]{mchardy+2004}
{McHardy} I.~M.,  {Papadakis} I.~E.,  {Uttley} P.,  {Page} M.~J.,   {Mason}
  K.~O.,  2004, \mn@doi [\mnras] {10.1111/j.1365-2966.2004.07376.x}, \href
  {http://adsabs.harvard.edu/abs/2004MNRAS.348..783M} {348, 783}

\bibitem[\protect\citeauthoryear{{McHardy}, {Koerding}, {Knigge}, {Uttley}  \&
  {Fender}}{{McHardy} et~al.}{2006}]{mchardy+2006}
{McHardy} I.~M.,  {Koerding} E.,  {Knigge} C.,  {Uttley} P.,   {Fender} R.~P.,
  2006, \mn@doi [\nat] {10.1038/nature05389}, \href
  {http://adsabs.harvard.edu/abs/2006Natur.444..730M} {444, 730}

\bibitem[\protect\citeauthoryear{{McHardy}, {Ar{\'e}valo}, {Uttley},
  {Papadakis}, {Summons}, {Brinkmann}  \& {Page}}{{McHardy}
  et~al.}{2007}]{mchardy+2007}
{McHardy} I.~M.,  {Ar{\'e}valo} P.,  {Uttley} P.,  {Papadakis} I.~E.,
  {Summons} D.~P.,  {Brinkmann} W.,   {Page} M.~J.,  2007, \mn@doi [\mnras]
  {10.1111/j.1365-2966.2007.12411.x}, \href
  {http://adsabs.harvard.edu/abs/2007MNRAS.382..985M} {382, 985}

\bibitem[\protect\citeauthoryear{{McKinney}, {Tchekhovskoy}  \&
  {Blandford}}{{McKinney} et~al.}{2012}]{mckinney+2012}
{McKinney} J.~C.,  {Tchekhovskoy} A.,   {Blandford} R.~D.,  2012, \mn@doi
  [\mnras] {10.1111/j.1365-2966.2012.21074.x}, \href
  {http://adsabs.harvard.edu/abs/2012MNRAS.423.3083M} {423, 3083}

\bibitem[\protect\citeauthoryear{{Merloni} \& {Fabian}}{{Merloni} \&
  {Fabian}}{2001}]{merloni_fabian}
{Merloni} A.,  {Fabian} A.~C.,  2001, \mn@doi [\mnras]
  {10.1046/j.1365-8711.2001.04925.x}, \href
  {http://adsabs.harvard.edu/abs/2001MNRAS.328..958M} {328, 958}

\bibitem[\protect\citeauthoryear{{Miyamoto} \& {Kitamoto}}{{Miyamoto} \&
  {Kitamoto}}{1989}]{miyamoto+89}
{Miyamoto} S.,  {Kitamoto} S.,  1989, \mn@doi [\nat] {10.1038/342773a0}, \href
  {http://adsabs.harvard.edu/abs/1989Natur.342..773M} {342, 773}

\bibitem[\protect\citeauthoryear{{Miyamoto}, {Kitamoto}, {Mitsuda}  \&
  {Dotani}}{{Miyamoto} et~al.}{1988}]{miyamoto+88}
{Miyamoto} S.,  {Kitamoto} S.,  {Mitsuda} K.,   {Dotani} T.,  1988, \mn@doi
  [\nat] {10.1038/336450a0}, \href
  {http://adsabs.harvard.edu/abs/1988Natur.336..450M} {336, 450}

\bibitem[\protect\citeauthoryear{{Nowak}, {Vaughan}, {Wilms}, {Dove}  \&
  {Begelman}}{{Nowak} et~al.}{1999}]{nowak+99}
{Nowak} M.~A.,  {Vaughan} B.~A.,  {Wilms} J.,  {Dove} J.~B.,   {Begelman}
  M.~C.,  1999, \mn@doi [\apj] {10.1086/306610}, \href
  {http://adsabs.harvard.edu/abs/1999ApJ...510..874N} {510, 874}

\bibitem[\protect\citeauthoryear{{Osterbrock} \& {Pogge}}{{Osterbrock} \&
  {Pogge}}{1987}]{osterbroke_pogge}
{Osterbrock} D.~E.,  {Pogge} R.~W.,  1987, \mn@doi [\apj] {10.1086/165810},
  \href {http://adsabs.harvard.edu/abs/1987ApJ...323..108O} {323, 108}

\bibitem[\protect\citeauthoryear{{Parker} et~al.,}{{Parker}
  et~al.}{2015}]{parker_pca}
{Parker} M.~L.,  et~al., 2015, \mn@doi [\mnras] {10.1093/mnras/stu2424}, \href
  {http://adsabs.harvard.edu/abs/2015MNRAS.447...72P} {447, 72}

\bibitem[\protect\citeauthoryear{{Ross} \& {Fabian}}{{Ross} \&
  {Fabian}}{2005}]{ross_fabian}
{Ross} R.~R.,  {Fabian} A.~C.,  2005, \mn@doi [\mnras]
  {10.1111/j.1365-2966.2005.08797.x}, \href
  {http://ukads.nottingham.ac.uk/abs/2005MNRAS.358..211R} {358, 211}

\bibitem[\protect\citeauthoryear{{Str{\"u}der} et~al.,}{{Str{\"u}der}
  et~al.}{2001}]{xmm_strueder}
{Str{\"u}der} L.,  et~al., 2001, \mn@doi [\aap] {10.1051/0004-6361:20000066},
  \href {http://adsabs.harvard.edu/abs/2001A\%26A...365L..18S} {365, L18}

\bibitem[\protect\citeauthoryear{{Uttley} \& {McHardy}}{{Uttley} \&
  {McHardy}}{2001}]{uttley+2001}
{Uttley} P.,  {McHardy} I.~M.,  2001, \mn@doi [\mnras]
  {10.1046/j.1365-8711.2001.04496.x}, \href
  {http://adsabs.harvard.edu/abs/2001MNRAS.323L..26U} {323, L26}

\bibitem[\protect\citeauthoryear{{Uttley}, {Cackett}, {Fabian}, {Kara}  \&
  {Wilkins}}{{Uttley} et~al.}{2014}]{reverb_review}
{Uttley} P.,  {Cackett} E.~M.,  {Fabian} A.~C.,  {Kara} E.,   {Wilkins} D.~R.,
  2014, \aapr

\bibitem[\protect\citeauthoryear{{Vaughan}, {Edelson}, {Warwick}  \&
  {Uttley}}{{Vaughan} et~al.}{2003}]{vaughan+2003}
{Vaughan} S.,  {Edelson} R.,  {Warwick} R.~S.,   {Uttley} P.,  2003, \mn@doi
  [\mnras] {10.1046/j.1365-2966.2003.07042.x}, \href
  {http://adsabs.harvard.edu/abs/2003MNRAS.345.1271V} {345, 1271}

\bibitem[\protect\citeauthoryear{{Vestergaard} \& {Peterson}}{{Vestergaard} \&
  {Peterson}}{2006}]{vestergaard+06}
{Vestergaard} M.,  {Peterson} B.~M.,  2006, \mn@doi [\apj] {10.1086/500572},
  \href {http://adsabs.harvard.edu/abs/2006ApJ...641..689V} {641, 689}

\bibitem[\protect\citeauthoryear{{Walton} et~al.,}{{Walton}
  et~al.}{2013}]{walton_hardlag}
{Walton} D.~J.,  et~al., 2013, \mn@doi [\apjl] {10.1088/2041-8205/777/2/L23},
  \href {http://adsabs.harvard.edu/abs/2013ApJ...777L..23W} {777, L23}

\bibitem[\protect\citeauthoryear{{Wilkins} \& {Fabian}}{{Wilkins} \&
  {Fabian}}{2011}]{1h0707_emis_paper}
{Wilkins} D.~R.,  {Fabian} A.~C.,  2011, \mn@doi [\mnras]
  {10.1111/j.1365-2966.2011.18458.x}, \href
  {http://adsabs.harvard.edu/abs/2011MNRAS.414.1269W} {414, 1269}

\bibitem[\protect\citeauthoryear{{Wilkins} \& {Fabian}}{{Wilkins} \&
  {Fabian}}{2012}]{understanding_emis_paper}
{Wilkins} D.~R.,  {Fabian} A.~C.,  2012, \mn@doi [\mnras]
  {10.1111/j.1365-2966.2012.21308.x}, \href
  {http://adsabs.harvard.edu/abs/2012MNRAS.424.1284W} {424, 1284}

\bibitem[\protect\citeauthoryear{{Wilkins} \& {Fabian}}{{Wilkins} \&
  {Fabian}}{2013}]{lag_spectra_paper}
{Wilkins} D.~R.,  {Fabian} A.~C.,  2013, \mn@doi [\mnras]
  {10.1093/mnras/sts591}, \href
  {http://adsabs.harvard.edu/abs/2013MNRAS.430..247W} {430, 247}

\bibitem[\protect\citeauthoryear{{Wilkins} \& {Gallo}}{{Wilkins} \&
  {Gallo}}{2015}]{mrk335_corona_paper}
{Wilkins} D.~R.,  {Gallo} L.~C.,  2015, \mn@doi [\mnras]
  {10.1093/mnras/stv162}, 449, 129

\bibitem[\protect\citeauthoryear{{Wilkins}, {Gallo}, {Grupe}, {Bonson},
  {Komossa}  \& {Fabian}}{{Wilkins} et~al.}{2015}]{mrk335_flare_paper}
{Wilkins} D.~R.,  {Gallo} L.~C.,  {Grupe} D.,  {Bonson} K.,  {Komossa} S.,
  {Fabian} A.~C.,  2015, \mn@doi [\mnras] {10.1093/mnras/stv2130}, \href
  {http://adsabs.harvard.edu/abs/2015MNRAS.454.4440W} {454, 4440}

\bibitem[\protect\citeauthoryear{{Wilkins}, {Cackett}, {Fabian}  \&
  {Reynolds}}{{Wilkins} et~al.}{2016}]{propagating_lag_paper}
{Wilkins} D.~R.,  {Cackett} E.~M.,  {Fabian} A.~C.,   {Reynolds} C.~S.,  2016,
  \mn@doi [\mnras] {10.1093/mnras/stw276}, \href
  {http://adsabs.harvard.edu/abs/2016MNRAS.458..200W} {458, 200}

\bibitem[\protect\citeauthoryear{{Wilkinson} \& {Uttley}}{{Wilkinson} \&
  {Uttley}}{2009}]{wilkinson_uttley}
{Wilkinson} T.,  {Uttley} P.,  2009, \mn@doi [\mnras]
  {10.1111/j.1365-2966.2009.15008.x}, \href
  {http://adsabs.harvard.edu/abs/2009MNRAS.397..666W} {397, 666}

\bibitem[\protect\citeauthoryear{{Zoghbi}, {Fabian}, {Reynolds}  \&
  {Cackett}}{{Zoghbi} et~al.}{2012}]{zoghbi+2012}
{Zoghbi} A.,  {Fabian} A.~C.,  {Reynolds} C.~S.,   {Cackett} E.~M.,  2012,
  \mn@doi [\mnras] {10.1111/j.1365-2966.2012.20587.x}, \href
  {http://adsabs.harvard.edu/abs/2012MNRAS.422..129Z} {422, 129}

\bibitem[\protect\citeauthoryear{{Zoghbi}, {Reynolds}, {Cackett}, {Miniutti},
  {Kara}  \& {Fabian}}{{Zoghbi} et~al.}{2013}]{zoghbi+2013}
{Zoghbi} A.,  {Reynolds} C.,  {Cackett} E.~M.,  {Miniutti} G.,  {Kara} E.,
  {Fabian} A.~C.,  2013, \mn@doi [\apj] {10.1088/0004-637X/767/2/121}, \href
  {http://adsabs.harvard.edu/abs/2013ApJ...767..121Z} {767, 121}

\makeatother
\end{thebibliography}

\label{lastpage}

\end{document}